\newif\ifjournal
\journalfalse

\ifjournal

\documentclass[useAMS,usenatbib,referee]{biom}

\else
\documentclass[letterpaper,12pt]{article}
\usepackage{geometry}
\usepackage{authblk}
\fi

\usepackage{multirow}
\usepackage{amsmath}
\usepackage{cases}
\usepackage{color}
\usepackage{rotating}
\usepackage{setspace}
\usepackage{graphicx}
\usepackage{subfigure}
\usepackage{float}
\usepackage{epsfig}
\usepackage{lscape}
\usepackage{multirow}
\usepackage{mathrsfs,amssymb}
\usepackage{lscape}
\usepackage{threeparttable}
\usepackage{array,booktabs}
\usepackage{mathtools,xparse}
\usepackage{rotating}
\usepackage{array}
\usepackage{algorithmic}
\usepackage{algorithm}

\newcolumntype{L}[1]{>{\raggedright\let\newline\\\arraybackslash\hspace{0pt}}m{#1}}
\newcolumntype{C}[1]{>{\centering\let\newline\\\arraybackslash\hspace{0pt}}m{#1}}
\newcolumntype{R}[1]{>{\raggedleft\let\newline\\\arraybackslash\hspace{0pt}}m{#1}}

\usepackage{times}
\usepackage{bm}
\usepackage{natbib}
\usepackage{graphicx,verbatim,array,multicol,fontenc,subfigure,pifont}
\usepackage{psfrag, fancybox, subfigure}
\usepackage{amsmath, amssymb, epsfig, color,mathrsfs,latexsym,mathrsfs}
\usepackage{pdfpages}
\usepackage{natbib}

\usepackage{array}
\newcolumntype{^}{>{\currentrowstyle}}

\usepackage{esvect}

\def\Oscr{\mathscr{O}}
\def\Dscr{\mathscr{D}}
\def\trans{^{\mbox{\scriptsize T}}}

\def\bX{\mathbf{X}}
\def\bS{\mathbf{S}}

\def\sumjp{\sum_{j=1}^p}

\def\Naive-Logistic{Y^*}
\def\bNaive-Logistic{\mathbf{Y}^*}

\def\bV{\mathbf{V}}

\def\Sschat{\widehat{\Ssc}}
\newcommand\ci{\perp\!\!\!\perp}

\def\G{{\bf G}}

\definecolor{darkred}{RGB}{150,50,50}
\definecolor{brown}{RGB}{250,100,100}
\definecolor{green}{RGB}{000,150,100}
\definecolor{purple}{RGB}{250,000,180}

\def\Lbb{\mathbb{L}}

\def\auc{\mbox{AUC}}
\def\roc{\mbox{ROC}}
\def\Ysc{\mathcal{Y}}
\def\Yschat{\widehat{\Ysc}}
\def\bSigma{\boldsymbol{\Sigma}}
\def\Esc{\mathcal{E}}

\def\bthetabar{\bar{\btheta}}
\def\bD{\mathbf{D}}
\def\bdetahat{\widehat{\boldsymbol{\eta}}}
\def\bdeta{\boldsymbol{\eta}}

\def\Omegahat{\widehat{\Omega}}
\def\bxi{\boldsymbol{\xi}}
\def\bzeta{\boldsymbol{\zeta}}

\def\Isc{\mathcal{I}}
\def\bfeta{\boldsymbol{\eta}}
\def\alphabar{\bar{\alpha}}

\def\bG{\widetilde{G}}
\def\bu{u}

\def\Ssc{\mathcal{S}}

\def\subpiy{_{\alpha,y}}
\def\subalphaty{_{\widehat\alpha,y}}
\def\subalphatzero{_{\widehat\alpha,0}}
\def\subalphatone{_{\widehat\alpha,1}}

\def\subalpbary{_{\alphabar,y}}

\def\subalpzero{_{\alpha,0}}
\def\subalpone{_{\alpha,1}}

\def\bbetabar{\bar{\bbeta}}

\def\phat{\widehat{p}}
\def\bvarphi{\boldsymbol{\varphi}}
\def\bpsi{\boldsymbol{\psi}}
\def\sumjp{\sum_{j=1}^p}
\def\U{\mathbf{U}}
\def\Pr{{\rm Pr}}
\def\EE{\mathbb{E}}

\newcommand\blfootnote[1]{%
  \begingroup
  \renewcommand\thefootnote{}\footnote{#1}%
  \addtocounter{footnote}{-1}%
  \endgroup
}

\ifjournal

\title[SCAN]{A Semiparametric Approach for Robust Modeling of Electronic Health Record Linked Biobank Data}

\author{Chuan Hong$^{1}$,
Molei Liu$^{2}$ and
Xinyi Wang$^{1,2*}$\email{}\\
$^{1}$Department of Biomedical Informatics, Harvard Medical School, Boston, MA, USA\\
$^{2}$Department of Biostatistics, Harvard T.H. Chan School of Public Health, Boston, MA, USA \\
}
\allowdisplaybreaks

\else

\title{A Semiparametric Approach for Robust and Efficient Learning with Biobank Data\blfootnote{The first two authors made equal contributions to this paper.}}

\author[1]{Molei Liu$^*$}
\author[2]{Xinyi Wang$^*$}
\author[3]{Chuan Hong}

\affil[1]{Department of Biostatistics, Columbia Mailman School of Public Health.}
\affil[2]{Department of Statistics, University of Chicago.}
\affil[3]{Department of Biostatistics and Bioinformatics, Duke University.}

\fi

\begin{document}
\setstretch{1.1}


%
%
%
%
%
\def\bzero{{\bf 0}}
\def\bone{{\bf 1}}
%
%
%
%
\def\ba{{\mbox{\boldmath$a$}}}
\def\bb{{\bf b}}
\def\bc{{\bf c}}
\def\bd{{\bf d}}
\def\be{{\bf e}}
\def\bdf{{\bf f}}
\def\bg{{\mbox{\boldmath$g$}}}
\def\bh{{\bf h}}
\def\bi{{\bf i}}
\def\bj{{\bf j}}
\def\bk{{\bf k}}
\def\bl{{\bf l}}
\def\bm{{\bf m}}
\def\bn{{\bf n}}
\def\bo{{\bf o}}
\def\bp{{\bf p}}
\def\bq{{\bf q}}
\def\br{{\bf r}}
\def\bs{{\bf s}}
\def\bt{{\bf t}}
\def\bu{{\bf u}}
\def\bv{{\bf v}}
\def\bw{{\bf w}}
\def\bx{{\bf x}}
\def\by{{\bf y}}
\def\bz{{\bf z}}
\def\bA{{\bf A}}
\def\bB{{\bf B}}
\def\bC{{\bf C}}
\def\bD{{\bf D}}
\def\bE{{\bf E}}
\def\bF{{\bf F}}
\def\bG{{\bf G}}
\def\bH{{\bf H}}
\def\bI{{\bf I}}
\def\bJ{{\bf J}}
\def\bK{{\bf K}}
\def\bL{{\bf L}}
\def\bM{{\bf M}}
\def\bN{{\bf N}}
\def\bO{{\bf O}}
\def\bP{{\bf P}}
\def\bQ{{\bf Q}}
\def\bR{{\bf R}}
\def\bS{{\bf S}}
\def\bT{{\bf T}}
\def\bU{{\bf U}}
\def\bV{{\bf V}}
\def\bW{{\bf W}}
\def\bX{{\bf X}}
\def\bY{{\bf Y}}
\def\bZ{{\bf Z}}
\def\smbZ{\scriptstyle{\bf Z}}
\def\smM{\scriptstyle{M}}
\def\smN{\scriptstyle{N}}
\def\smbT{\scriptstyle{\bf T}}
%
%
%
%
\def\thick#1{\hbox{\rlap{$#1$}\kern0.25pt\rlap{$#1$}\kern0.25pt$#1$}}
\def\balpha{\boldsymbol{\alpha}}
\def\bbeta{\boldsymbol{\beta}}
\def\bgamma{\boldsymbol{\gamma}}
\def\bdelta{\boldsymbol{\delta}}
\def\bepsilon{\boldsymbol{\epsilon}}
\def\bvarepsilon{\boldsymbol{\varepsilon}}
\def\bzeta{\boldsymbol{\zeta}}
\def\bdeta{\boldsymbol{\eta}}
\def\btheta{\boldsymbol{\theta}}
\def\biota{\boldsymbol{\iota}}
\def\bkappa{\boldsymbol{\kappa}}
\def\blambda{\boldsymbol{\lambda}}
\def\bmu{\boldsymbol{\mu}}
\def\bnu{\boldsymbol{\nu}}
\def\bxi{\boldsymbol{\xi}}
\def\bomicron{\boldsymbol{\omicron}}
\def\bpi{\boldsymbol{\pi}}
\def\brho{\boldsymbol{\rho}}
\def\bsigma{\boldsymbol{\sigma}}
\def\btau{\boldsymbol{\tau}}
\def\bupsilon{\boldsymbol{\upsilon}}
\def\bphi{\boldsymbol{\phi}}
\def\bchi{\boldsymbol{\chi}}
\def\bpsi{\boldsymbol{\psi}}
\def\bomega{\boldsymbol{\omega}}
\def\bAlpha{\boldsymbol{\Alpha}}
\def\bBeta{\boldsymbol{\Beta}}
\def\bGamma{\boldsymbol{\Gamma}}
\def\bDelta{\boldsymbol{\Delta}}
\def\bEpsilon{\boldsymbol{\Epsilon}}
\def\bZeta{\boldsymbol{\Zeta}}
\def\bEta{\boldsymbol{\Eta}}
\def\bTheta{\boldsymbol{\Theta}}
\def\bIota{\boldsymbol{\Iota}}
\def\bKappa{\boldsymbol{\Kappa}}
\def\bLambda{{\boldsymbol{\Lambda}}}
\def\bMu{\boldsymbol{\Mu}}
\def\bNu{\boldsymbol{\Nu}}
\def\bXi{\boldsymbol{\Xi}}
\def\bOmicron{\boldsymbol{\Omicron}}
\def\bPi{\boldsymbol{\Pi}}
\def\bRho{\boldsymbol{\Rho}}
\def\bSigma{\boldsymbol{\Sigma}}
\def\bTau{\boldsymbol{\Tau}}
\def\bUpsilon{\boldsymbol{\Upsilon}}
\def\bPhi{\boldsymbol{\Phi}}
\def\bChi{\boldsymbol{\Chi}}
\def\bPsi{\boldsymbol{\Psi}}
\def\bOmega{\boldsymbol{\Omega}}
%
%
%
\def\smalpha{{{\scriptstyle{\alpha}}}}
\def\smbeta{{{\scriptstyle{\beta}}}}
\def\smgamma{{{\scriptstyle{\gamma}}}}
\def\smdelta{{{\scriptstyle{\delta}}}}
\def\smepsilon{{{\scriptstyle{\epsilon}}}}
\def\smvarepsilon{{{\scriptstyle{\varepsilon}}}}
\def\smzeta{{{\scriptstyle{\zeta}}}}
\def\smdeta{{{\scriptstyle{\eta}}}}
\def\smtheta{{{\scriptstyle{\theta}}}}
\def\smiota{{{\scriptstyle{\iota}}}}
\def\smkappa{{{\scriptstyle{\kappa}}}}
\def\smlambda{{{\scriptstyle{\lambda}}}}
\def\smmu{{{\scriptstyle{\mu}}}}
\def\smnu{{{\scriptstyle{\nu}}}}
\def\smxi{{{\scriptstyle{\xi}}}}
\def\smomicron{{{\scriptstyle{\omicron}}}}
\def\smpi{{{\scriptstyle{\pi}}}}
\def\smrho{{{\scriptstyle{\rho}}}}
\def\smsigma{{{\scriptstyle{\sigma}}}}
\def\smtau{{{\scriptstyle{\tau}}}}
\def\smupsilon{{{\scriptstyle{\upsilon}}}}
\def\smphi{{{\scriptstyle{\phi}}}}
\def\smchi{{{\scriptstyle{\chi}}}}
\def\smpsi{{{\scriptstyle{\psi}}}}
\def\smomega{{{\scriptstyle{\omega}}}}
\def\smAlpha{{{\scriptstyle{\Alpha}}}}
\def\smBeta{{{\scriptstyle{\Beta}}}}
\def\smGamma{{{\scriptstyle{\Gamma}}}}
\def\smDelta{{{\scriptstyle{\Delta}}}}
\def\smEpsilon{{{\scriptstyle{\Epsilon}}}}
\def\smZeta{{{\scriptstyle{\Zeta}}}}
\def\smEta{{{\scriptstyle{\Eta}}}}
\def\smTheta{{{\scriptstyle{\Theta}}}}
\def\smIota{{{\scriptstyle{\Iota}}}}
\def\smKappa{{{\scriptstyle{\Kappa}}}}
\def\smLambda{{{\scriptstyle{\Lambda}}}}
\def\smMu{{{\scriptstyle{\Mu}}}}
\def\smNu{{{\scriptstyle{\Nu}}}}
\def\smXi{{{\scriptstyle{\Xi}}}}
\def\smOmicron{{{\scriptstyle{\Omicron}}}}
\def\smPi{{{\scriptstyle{\Pi}}}}
\def\smRho{{{\scriptstyle{\Rho}}}}
\def\smSigma{{{\scriptstyle{\Sigma}}}}
\def\smTau{{{\scriptstyle{\Tau}}}}
\def\smUpsilon{{{\scriptstyle{\Upsilon}}}}
\def\smPhi{{{\scriptstyle{\Phi}}}}
\def\smChi{{{\scriptstyle{\Chi}}}}
\def\smPsi{{{\scriptstyle{\Psi}}}}
\def\smOmega{{{\scriptstyle{\Omega}}}}
%
%

%
\def\smbalpha{\boldsymbol{{\scriptstyle{\alpha}}}}
\def\smbbeta{\boldsymbol{{\scriptstyle{\beta}}}}
\def\smbgamma{\boldsymbol{{\scriptstyle{\gamma}}}}
\def\smbdelta{\boldsymbol{{\scriptstyle{\delta}}}}
\def\smbepsilon{\boldsymbol{{\scriptstyle{\epsilon}}}}
\def\smbvarepsilon{\boldsymbol{{\scriptstyle{\varepsilon}}}}
\def\smbzeta{\boldsymbol{{\scriptstyle{\zeta}}}}
\def\smbdeta{\boldsymbol{{\scriptstyle{\eta}}}}
\def\smbtheta{\boldsymbol{{\scriptstyle{\theta}}}}
\def\smbiota{\boldsymbol{{\scriptstyle{\iota}}}}
\def\smbkappa{\boldsymbol{{\scriptstyle{\kappa}}}}
\def\smblambda{\boldsymbol{{\scriptstyle{\lambda}}}}
\def\smbmu{\boldsymbol{{\scriptstyle{\mu}}}}
\def\smbnu{\boldsymbol{{\scriptstyle{\nu}}}}
\def\smbxi{\boldsymbol{{\scriptstyle{\xi}}}}
\def\smbomicron{\boldsymbol{{\scriptstyle{\omicron}}}}
\def\smbpi{\boldsymbol{{\scriptstyle{\pi}}}}
\def\smbrho{\boldsymbol{{\scriptstyle{\rho}}}}
\def\smbsigma{\boldsymbol{{\scriptstyle{\sigma}}}}
\def\smbtau{\boldsymbol{{\scriptstyle{\tau}}}}
\def\smbupsilon{\boldsymbol{{\scriptstyle{\upsilon}}}}
\def\smbphi{\boldsymbol{{\scriptstyle{\phi}}}}
\def\smbchi{\boldsymbol{{\scriptstyle{\chi}}}}
\def\smbpsi{\boldsymbol{{\scriptstyle{\psi}}}}
\def\smbomega{\boldsymbol{{\scriptstyle{\omega}}}}
\def\smbAlpha{\boldsymbol{{\scriptstyle{\Alpha}}}}
\def\smbBeta{\boldsymbol{{\scriptstyle{\Beta}}}}
\def\smbGamma{\boldsymbol{{\scriptstyle{\Gamma}}}}
\def\smbDelta{\boldsymbol{{\scriptstyle{\Delta}}}}
\def\smbEpsilon{\boldsymbol{{\scriptstyle{\Epsilon}}}}
\def\smbZeta{\boldsymbol{{\scriptstyle{\Zeta}}}}
\def\smbEta{\boldsymbol{{\scriptstyle{\Eta}}}}
\def\smbTheta{\boldsymbol{{\scriptstyle{\Theta}}}}
\def\smbIota{\boldsymbol{{\scriptstyle{\Iota}}}}
\def\smbKappa{\boldsymbol{{\scriptstyle{\Kappa}}}}
\def\smbLambda{\boldsymbol{{\scriptstyle{\Lambda}}}}
\def\smbMu{\boldsymbol{{\scriptstyle{\Mu}}}}
\def\smbNu{\boldsymbol{{\scriptstyle{\Nu}}}}
\def\smbXi{\boldsymbol{{\scriptstyle{\Xi}}}}
\def\smbOmicron{\boldsymbol{{\scriptstyle{\Omicron}}}}
\def\smbPi{\boldsymbol{{\scriptstyle{\Pi}}}}
\def\smbRho{\boldsymbol{{\scriptstyle{\Rho}}}}
\def\smbSigma{\boldsymbol{{\scriptstyle{\Sigma}}}}
\def\smbTau{\boldsymbol{{\scriptstyle{\Tau}}}}
\def\smbUpsilon{\boldsymbol{{\scriptstyle{\Upsilon}}}}
\def\smbPhi{\boldsymbol{{\scriptstyle{\Phi}}}}
\def\smbChi{\boldsymbol{{\scriptstyle{\Chi}}}}
\def\smbPsi{\boldsymbol{{\scriptstyle{\Psi}}}}
\def\smbOmega{\boldsymbol{{\scriptstyle{\Omega}}}}
%
%
%
%
\def\ahat{{\widehat a}}
\def\bhat{{\widehat b}}
\def\chat{{\widehat c}}
\def\dhat{{\widehat d}}
\def\ehat{{\widehat e}}
\def\fhat{{\widehat f}}
\def\ghat{{\widehat g}}
\def\hhat{{\widehat h}}
\def\ihat{{\widehat i}}
\def\jhat{{\widehat j}}
\def\khat{{\widehat k}}
\def\lhat{{\widehat l}}
\def\mhat{{\widehat m}}
\def\nhat{{\widehat n}}
\def\ohat{{\widehat o}}
\def\phat{{\widehat p}}
\def\qhat{{\widehat q}}
\def\rhat{{\widehat r}}
\def\shat{{\widehat s}}
\def\that{{\widehat t}}
\def\uhat{{\widehat u}}
\def\vhat{{\widehat v}}
\def\what{{\widehat w}}
\def\xhat{{\widehat x}}
\def\yhat{{\widehat y}}
\def\zhat{{\widehat z}}
\def\Ahat{{\widehat A}}
\def\Bhat{{\widehat B}}
\def\Chat{{\widehat C}}
\def\Dhat{{\widehat D}}
\def\Ehat{{\widehat E}}
\def\Fhat{{\widehat F}}
\def\Ghat{{\widehat G}}
\def\Hhat{{\widehat H}}
\def\Ihat{{\widehat I}}
\def\Jhat{{\widehat J}}
\def\Khat{{\widehat K}}
\def\Lhat{{\widehat L}}
\def\Mhat{{\widehat M}}
\def\Nhat{{\widehat N}}
\def\Ohat{{\widehat O}}
\def\Phat{{\widehat P}}
\def\Qhat{{\widehat Q}}
\def\Rhat{{\widehat R}}
\def\Shat{{\widehat S}}
\def\That{{\widehat T}}
\def\Uhat{{\widehat U}}
\def\Vhat{{\widehat V}}
\def\What{{\widehat W}}
\def\Xhat{{\widehat X}}
\def\Yhat{{\widehat Y}}
\def\Zhat{{\widehat Z}}
%
%
%
\def\atilde{{\widetilde a}}
\def\btilde{{\widetilde b}}
\def\ctilde{{\widetilde c}}
\def\dtilde{{\widetilde d}}
\def\etilde{{\widetilde e}}
\def\ftilde{{\widetilde f}}
\def\gtilde{{\widetilde g}}
\def\htilde{{\widetilde h}}
\def\itilde{{\widetilde i}}
\def\jtilde{{\widetilde j}}
\def\ktilde{{\widetilde k}}
\def\ltilde{{\widetilde l}}
\def\mtilde{{\widetilde m}}
\def\ntilde{{\widetilde n}}
\def\otilde{{\widetilde o}}
\def\ptilde{{\widetilde p}}
\def\qtilde{{\widetilde q}}
\def\rtilde{{\widetilde r}}
\def\stilde{{\widetilde s}}
\def\ttilde{{\widetilde t}}
\def\utilde{{\widetilde u}}
\def\vtilde{{\widetilde v}}
\def\wtilde{{\widetilde w}}
\def\xtilde{{\widetilde x}}
\def\ytilde{{\widetilde y}}
\def\ztilde{{\widetilde z}}
\def\Atilde{{\widetilde A}}
\def\Btilde{{\widetilde B}}
\def\Ctilde{{\widetilde C}}
\def\Dtilde{{\widetilde D}}
\def\Etilde{{\widetilde E}}
\def\Ftilde{{\widetilde F}}
\def\Gtilde{{\widetilde G}}
\def\Htilde{{\widetilde H}}
\def\Itilde{{\widetilde I}}
\def\Jtilde{{\widetilde J}}
\def\Ktilde{{\widetilde K}}
\def\Ltilde{{\widetilde L}}
\def\Mtilde{{\widetilde M}}
\def\Ntilde{{\widetilde N}}
\def\Otilde{{\widetilde O}}
\def\Ptilde{{\widetilde P}}
\def\Qtilde{{\widetilde Q}}
\def\Rtilde{{\widetilde R}}
\def\Stilde{{\widetilde S}}
\def\Ttilde{{\widetilde T}}
\def\Utilde{{\widetilde U}}
\def\Vtilde{{\widetilde V}}
\def\Wtilde{{\widetilde W}}
\def\Xtilde{{\widetilde X}}
\def\Ytilde{{\widetilde Y}}
\def\Ztilde{{\widetilde Z}}
%
%
%
%
\def\bahat{{\widehat \ba}}
\def\bbhat{{\widehat \bb}}
\def\bchat{{\widehat \bc}}
\def\bdhat{{\widehat \bd}}
\def\behat{{\widehat \be}}
\def\bfhat{{\widehat \bf}}
\def\bghat{{\widehat \bg}}
\def\bhhat{{\widehat \bh}}
\def\bihat{{\widehat \bi}}
\def\bjhat{{\widehat \bj}}
\def\bkhat{{\widehat \bk}}
\def\blhat{{\widehat \bl}}
\def\bmhat{{\widehat \bm}}
\def\bnhat{{\widehat \bn}}
\def\bohat{{\widehat \bo}}
\def\bphat{{\widehat \bp}}
\def\bqhat{{\widehat \bq}}
\def\brhat{{\widehat \br}}
\def\bshat{{\widehat \bs}}
\def\bthat{{\widehat \bt}}
\def\buhat{{\widehat \bu}}
\def\bvhat{{\widehat \bv}}
\def\bwhat{{\widehat \bw}}
\def\bxhat{{\widehat \bx}}
\def\byhat{{\widehat \by}}
\def\bzhat{{\widehat \bz}}
\def\bAhat{{\widehat \bA}}
\def\bBhat{{\widehat \bB}}
\def\bChat{{\widehat \bC}}
\def\bDhat{{\widehat \bD}}
\def\bEhat{{\widehat \bE}}
\def\bFhat{{\widehat \bF}}
\def\bGhat{{\widehat \bG}}
\def\bHhat{{\widehat \bH}}
\def\bIhat{{\widehat \bI}}
\def\bJhat{{\widehat \bJ}}
\def\bKhat{{\widehat \bK}}
\def\bLhat{{\widehat \bL}}
\def\bMhat{{\widehat \bM}}
\def\bNhat{{\widehat \bN}}
\def\bOhat{{\widehat \bO}}
\def\bPhat{{\widehat \bP}}
\def\bQhat{{\widehat \bQ}}
\def\bRhat{{\widehat \bR}}
\def\bShat{{\widehat \bS}}
\def\bThat{{\widehat \bT}}
\def\bUhat{{\widehat \bU}}
\def\bVhat{{\widehat \bV}}
\def\bWhat{{\widehat \bW}}
\def\bXhat{{\widehat \bX}}
\def\bYhat{{\widehat \bY}}
\def\bZhat{{\widehat \bZ}}
%
%
%
%
%
\def\batilde{{\widetilde \ba}}
\def\bbtilde{{\widetilde \bb}}
\def\bctilde{{\widetilde \bc}}
\def\bdtilde{{\widetilde \bd}}
\def\betilde{{\widetilde \be}}
\def\bftilde{{\widetilde \bf}}
\def\bgtilde{{\widetilde \bg}}
\def\bhtilde{{\widetilde \bh}}
\def\bitilde{{\widetilde \bi}}
\def\bjtilde{{\widetilde \bj}}
\def\bktilde{{\widetilde \bk}}
\def\bltilde{{\widetilde \bl}}
\def\bmtilde{{\widetilde \bm}}
\def\bntilde{{\widetilde \bn}}
\def\botilde{{\widetilde \bo}}
\def\bptilde{{\widetilde \bp}}
\def\bqtilde{{\widetilde \bq}}
\def\brtilde{{\widetilde \br}}
\def\bstilde{{\widetilde \bs}}
\def\bttilde{{\widetilde \bt}}
\def\butilde{{\widetilde \bu}}
\def\bvtilde{{\widetilde \bv}}
\def\bwtilde{{\widetilde \bw}}
\def\bxtilde{{\widetilde \bx}}
\def\bytilde{{\widetilde \by}}
\def\bztilde{{\widetilde \bz}}
\def\bAtilde{{\widetilde \bA}}
\def\bBtilde{{\widetilde \bB}}
\def\bCtilde{{\widetilde \bC}}
\def\bDtilde{{\widetilde \bD}}
\def\bEtilde{{\widetilde \bE}}
\def\bFtilde{{\widetilde \bF}}
\def\bGtilde{{\widetilde \bG}}
\def\bHtilde{{\widetilde \bH}}
\def\bItilde{{\widetilde \bI}}
\def\bJtilde{{\widetilde \bJ}}
\def\bKtilde{{\widetilde \bK}}
\def\bLtilde{{\widetilde \bL}}
\def\bMtilde{{\widetilde \bM}}
\def\bNtilde{{\widetilde \bN}}
\def\bOtilde{{\widetilde \bO}}
\def\bPtilde{{\widetilde \bP}}
\def\bQtilde{{\widetilde \bQ}}
\def\bRtilde{{\widetilde \bR}}
\def\bStilde{{\widetilde \bS}}
\def\bTtilde{{\widetilde \bT}}
\def\bUtilde{{\widetilde \bU}}
\def\bVtilde{{\widetilde \bV}}
\def\bWtilde{{\widetilde \bW}}
\def\bXtilde{{\widetilde \bX}}
\def\bYtilde{{\widetilde \bY}}
\def\bZtilde{{\widetilde \bZ}}
%
%
%
%
%
%
\def\alphahat{{\widehat\alpha}}
\def\betahat{{\widehat\beta}}
\def\gammahat{{\widehat\gamma}}
\def\deltahat{{\widehat\delta}}
\def\epsilonhat{{\widehat\epsilon}}
\def\varepsilonhat{{\widehat\varepsilon}}
\def\zetahat{{\widehat\zeta}}
\def\etahat{{\widehat\eta}}
\def\thetahat{{\widehat\theta}}
\def\iotahat{{\widehat\iota}}
\def\kappahat{{\widehat\kappa}}
\def\lambdahat{{\widehat\lambda}}
\def\muhat{{\widehat\mu}}
\def\nuhat{{\widehat\nu}}
\def\xihat{{\widehat\xi}}
\def\omicronhat{{\widehat\omicron}}
\def\pihat{{\widehat\pi}}
\def\rhohat{{\widehat\rho}}
\def\sigmahat{{\widehat\sigma}}
\def\tauhat{{\widehat\tau}}
\def\upsilonhat{{\widehat\upsilon}}
\def\phihat{{\widehat\phi}}
\def\chihat{{\widehat\chi}}
\def\psihat{{\widehat\psi}}
\def\omegahat{{\widehat\omega}}
\def\Alphahat{{\widehat\Alpha}}
\def\Betahat{{\widehat\Beta}}
\def\Gammahat{{\widehat\Gamma}}
\def\Deltahat{{\widehat\Delta}}
\def\Epsilonhat{{\widehat\Epsilon}}
\def\Zetahat{{\widehat\Zeta}}
\def\Etahat{{\widehat\Eta}}
\def\Thetahat{{\widehat\Theta}}
\def\Iotahat{{\widehat\Iota}}
\def\Kappahat{{\widehat\Kappa}}
\def\Lambdahat{{\widehat\Lambda}}
\def\Muhat{{\widehat\Mu}}
\def\Nuhat{{\widehat\Nu}}
\def\Xihat{{\widehat\Xi}}
\def\Omicronhat{{\widehat\Omicron}}
\def\Pihat{{\widehat\Pi}}
\def\Rhohat{{\widehat\Rho}}
\def\Sigmahat{{\widehat\Sigma}}
\def\Tauhat{{\widehat\Tau}}
\def\Upsilonhat{{\widehat\Upsilon}}
\def\Phihat{{\widehat\Phi}}
\def\Chihat{{\widehat\Chi}}
\def\Psihat{{\widehat\Psi}}
\def\Omegahat{{\widehat\Omega}}
%
%
%
%
%
\def\alphatilde{{\widetilde\alpha}}
\def\betatilde{{\widetilde\beta}}
\def\gammatilde{{\widetilde\gamma}}
\def\deltatilde{{\widetilde\delta}}
\def\epsilontilde{{\widetilde\epsilon}}
\def\varepsilontilde{{\widetilde\varepsilon}}
\def\zetatilde{{\widetilde\zeta}}
\def\etatilde{{\widetilde\eta}}
\def\thetatilde{{\widetilde\theta}}
\def\iotatilde{{\widetilde\iota}}
\def\kappatilde{{\widetilde\kappa}}
\def\lambdatilde{{\widetilde\lambda}}
\def\mutilde{{\widetilde\mu}}
\def\nutilde{{\widetilde\nu}}
\def\xitilde{{\widetilde\xi}}
\def\omicrontilde{{\widetilde\omicron}}
\def\pitilde{{\widetilde\pi}}
\def\rhotilde{{\widetilde\rho}}
\def\sigmatilde{{\widetilde\sigma}}
\def\tautilde{{\widetilde\tau}}
\def\upsilontilde{{\widetilde\upsilon}}
\def\phitilde{{\widetilde\phi}}
\def\chitilde{{\widetilde\chi}}
\def\psitilde{{\widetilde\psi}}
\def\omegatilde{{\widetilde\omega}}
\def\Alphatilde{{\widetilde\Alpha}}
\def\Betatilde{{\widetilde\Beta}}
\def\Gammatilde{{\widetilde\Gamma}}
\def\Deltatilde{{\widetilde\Delta}}
\def\Epsilontilde{{\widetilde\Epsilon}}
\def\Zetatilde{{\widetilde\Zeta}}
\def\Etatilde{{\widetilde\Eta}}
\def\Thetatilde{{\widetilde\Theta}}
\def\Iotatilde{{\widetilde\Iota}}
\def\Kappatilde{{\widetilde\Kappa}}
\def\Lambdatilde{{\widetilde\Lambda}}
\def\Mutilde{{\widetilde\Mu}}
\def\Nutilde{{\widetilde\Nu}}
\def\Xitilde{{\widetilde\Xi}}
\def\Omicrontilde{{\widetilde\Omicron}}
\def\Pitilde{{\widetilde\Pi}}
\def\Rhotilde{{\widetilde\Rho}}
\def\Sigmatilde{{\widetilde\Sigma}}
\def\Tautilde{{\widetilde\Tau}}
\def\Upsilontilde{{\widetilde\Upsilon}}
\def\Phitilde{{\widetilde\Phi}}
\def\Chitilde{{\widetilde\Chi}}
\def\Psitilde{{\widetilde\Psi}}
\def\Omegatilde{{\widetilde\Omega}}
%
%
%
%
%
%
\def\balphahat{{\widehat\balpha}}
\def\bbetahat{{\widehat\bbeta}}
\def\bgammahat{{\widehat\bgamma}}
\def\bdeltahat{{\widehat\bdelta}}
\def\bepsilonhat{{\widehat\bepsilon}}
\def\bzetahat{{\widehat\bzeta}}
\def\bdetahat{{\widehat\bdeta}}
\def\bthetahat{{\widehat\btheta}}
\def\biotahat{{\widehat\biota}}
\def\bkappahat{{\widehat\bkappa}}
\def\blambdahat{{\widehat\blambda}}
\def\bmuhat{{\widehat\bmu}}
\def\bnuhat{{\widehat\bnu}}
\def\bxihat{{\widehat\bxi}}
\def\bomicronhat{{\widehat\bomicron}}
\def\bpihat{{\widehat\bpi}}
\def\brhohat{{\widehat\brho}}
\def\bsigmahat{{\widehat\bsigma}}
\def\btauhat{{\widehat\btau}}
\def\bupsilonhat{{\widehat\bupsilon}}
\def\bphihat{{\widehat\bphi}}
\def\bchihat{{\widehat\bchi}}
\def\bpsihat{{\widehat\bpsi}}
\def\bomegahat{{\widehat\bomega}}
\def\bAlphahat{{\widehat\bAlpha}}
\def\bBetahat{{\widehat\bBeta}}
\def\bGammahat{{\widehat\bGamma}}
\def\bDeltahat{{\widehat\bDelta}}
\def\bEpsilonhat{{\widehat\bEpsilon}}
\def\bZetahat{{\widehat\bZeta}}
\def\bEtahat{{\widehat\bEta}}
\def\bThetahat{{\widehat\bTheta}}
\def\bIotahat{{\widehat\bIota}}
\def\bKappahat{{\widehat\bKappa}}
\def\bLambdahat{{\widehat\bLambda}}
\def\bMuhat{{\widehat\bMu}}
\def\bNuhat{{\widehat\bNu}}
\def\bXihat{{\widehat\bXi}}
\def\bOmicronhat{{\widehat\bOmicron}}
\def\bPihat{{\widehat\bPi}}
\def\bRhohat{{\widehat\bRho}}
\def\bSigmahat{{\widehat\bSigma}}
\def\bTauhat{{\widehat\bTau}}
\def\bUpsilonhat{{\widehat\bUpsilon}}
\def\bPhihat{{\widehat\bPhi}}
\def\bChihat{{\widehat\bChi}}
\def\bPsihat{{\widehat\bPsi}}
\def\bOmegahat{{\widehat\bOmega}}%
\def\balphahattrans{{\balphahat^{_{\transpose}}}}
\def\bbetahattrans{{\bbetahat^{_{\transpose}}}}
\def\bgammahattrans{{\bgammahat^{_{\transpose}}}}
\def\bdeltahattrans{{\bdeltahat^{_{\transpose}}}}
\def\bepsilonhattrans{{\bepsilonhat^{_{\transpose}}}}
\def\bzetahattrans{{\bzetahat^{_{\transpose}}}}
\def\bdetahattrans{{\bdetahat^{_{\transpose}}}}
\def\bthetahattrans{{\bthetahat^{_{\transpose}}}}
\def\biotahattrans{{\biotahat^{_{\transpose}}}}
\def\bkappahattrans{{\bkappahat^{_{\transpose}}}}
\def\blambdahattrans{{\blambdahat^{_{\transpose}}}}
\def\bmuhattrans{{\bmuhat^{_{\transpose}}}}
\def\bnuhattrans{{\bnuhat^{_{\transpose}}}}
\def\bxihattrans{{\bxihat^{_{\transpose}}}}
\def\bomicronhattrans{{\bomicronhat^{_{\transpose}}}}
\def\bpihattrans{{\bpihat^{_{\transpose}}}}
\def\brhohattrans{{\brhohat^{_{\transpose}}}}
\def\bsigmahattrans{{\bsigmahat^{_{\transpose}}}}
\def\btauhattrans{{\btauhat^{_{\transpose}}}}
\def\bupsilonhattrans{{\bupsilonhat^{_{\transpose}}}}
\def\bphihattrans{{\bphihat^{_{\transpose}}}}
\def\bchihattrans{{\bchihat^{_{\transpose}}}}
\def\bpsihattrans{{\bpsihat^{_{\transpose}}}}
\def\bomegahattrans{{\bomegahat^{_{\transpose}}}}
\def\bAlphahattrans{{\bAlphahat^{_{\transpose}}}}
\def\bBetahattrans{{\bBetahat^{_{\transpose}}}}
\def\bGammahattrans{{\bGammahat^{_{\transpose}}}}
\def\bDeltahattrans{{\bDeltahat^{_{\transpose}}}}
\def\bEpsilonhattrans{{\bEpsilonhat^{_{\transpose}}}}
\def\bZetahattrans{{\bZetahat^{_{\transpose}}}}
\def\bEtahattrans{{\bEtahat^{_{\transpose}}}}
\def\bThetahattrans{{\bThetahat^{_{\transpose}}}}
\def\bIotahattrans{{\bIotahat^{_{\transpose}}}}
\def\bKappahattrans{{\bKappahat^{_{\transpose}}}}
\def\bLambdahattrans{{\bLambdahat^{_{\transpose}}}}
\def\bMuhattrans{{\bMuhat^{_{\transpose}}}}
\def\bNuhattrans{{\bNuhat^{_{\transpose}}}}
\def\bXihattrans{{\bXihat^{_{\transpose}}}}
\def\bOmicronhattrans{{\bOmicronhat^{_{\transpose}}}}
\def\bPihattrans{{\bPihat^{_{\transpose}}}}
\def\bRhohattrans{{\bRhohat^{_{\transpose}}}}
\def\bSigmahattrans{{\bSigmahat^{_{\transpose}}}}
\def\bTauhattrans{{\bTauhat^{_{\transpose}}}}
\def\bUpsilonhattrans{{\bUpsilonhat^{_{\transpose}}}}
\def\bPhihattrans{{\bPhihat^{_{\transpose}}}}
\def\bChihattrans{{\bChihat^{_{\transpose}}}}
\def\bPsihattrans{{\bPsihat^{_{\transpose}}}}
\def\bOmegahattrans{{\bOmegahat^{_{\transpose}}}}%
%
\def\smbalpha{\widehat{\smbalpha}}
\def\smbbetahat{\widehat{\smbbeta}}
\def\smbgammahat{\widehat{\smbgamma}}
\def\smbdeltahat{\widehat{\smbdelta}}
\def\smbepsilonhat{\widehat{\smbepsilon}}
\def\smbvarepsilonhat{\widehat{\smbvarepsilon}}
\def\smbzetahat{\widehat{\smbzeta}}
\def\smbdetahat{\widehat{\smbeta}}
\def\smbthetahat{\widehat{\smbtheta}}
\def\smbiotahat{\widehat{\smbiota}}
\def\smbkappahat{\widehat{\smbkappa}}
\def\smblambdahat{\widehat{\smblambda}}
\def\smbmuhat{\widehat{\smbmu}}
\def\smbnuhat{\widehat{\smbnu}}
\def\smbxihat{\widehat{\smbxi}}
\def\smbomicronhat{\widehat{\smbomicron}}
\def\smbpihat{\widehat{\smbpi}}
\def\smbrhohat{\widehat{\smbrho}}
\def\smbsigmahat{\widehat{\smbsigma}}
\def\smbtauhat{\widehat{\smbtau}}
\def\smbupsilonhat{\widehat{\smbupsilon}}
\def\smbphihat{\widehat{\smbphi}}
\def\smbchihat{\widehat{\smbchi}}
\def\smbpsihat{\widehat{\smbpsi}}
\def\smbomegahat{\widehat{\smbomega}}
\def\smbAlphahat{\widehat{\smbAlpha}}
\def\smbBetahat{\widehat{\smbBeta}}
\def\smbGammahat{\widehat{\smbGamma}}
\def\smbDeltahat{\widehat{\smbDelta}}
\def\smbEpsilonhat{\widehat{\smbEpsilon}}
\def\smbZetahat{\widehat{\smbZeta}}
\def\smbEtahat{\widehat{\smbEta}}
\def\smbThetahat{\widehat{\smbTheta}}
\def\smbIotahat{\widehat{\smbIota}}
\def\smbKappahat{\widehat{\smbKappa}}
\def\smbLambdahat{\widehat{\smbLambda}}
\def\smbMuhat{\widehat{\smbMu}}
\def\smbNuhat{\widehat{\smbNu}}
\def\smbXihat{\widehat{\smbXi}}
\def\smbOmicronhat{\widehat{\smbOmicron}}
\def\smbPihat{\widehat{\smbPi}}
\def\smbRhohat{\widehat{\smbRho}}
\def\smbSigmahat{\widehat{\smbSigma}}
\def\smbTauhat{\widehat{\smbTau}}
\def\smbUpsilonhat{\widehat{\smbUpsilon}}
\def\smbPhihat{\widehat{\smbPhi}}
\def\smbChihat{\widehat{\smbChi}}
\def\smbPsihat{\widehat{\smbPsi}}
\def\smbOmegahat{\widehat{\smbOmega}}
%
%
%
%
%
\def\balphatilde{{\widetilde\balpha}}
\def\bbetatilde{{\widetilde\bbeta}}
\def\bgammatilde{{\widetilde\bgamma}}
\def\bdeltatilde{{\widetilde\bdelta}}
\def\bepsilontilde{{\widetilde\bepsilon}}
\def\bzetatilde{{\widetilde\bzeta}}
\def\bdetatilde{{\widetilde\bdeta}}
\def\bthetatilde{{\widetilde\btheta}}
\def\biotatilde{{\widetilde\biota}}
\def\bkappatilde{{\widetilde\bkappa}}
\def\blambdatilde{{\widetilde\blambda}}
\def\bmutilde{{\widetilde\bmu}}
\def\bnutilde{{\widetilde\bnu}}
\def\bxitilde{{\widetilde\bxi}}
\def\bomicrontilde{{\widetilde\bomicron}}
\def\bpitilde{{\widetilde\bpi}}
\def\brhotilde{{\widetilde\brho}}
\def\bsigmatilde{{\widetilde\bsigma}}
\def\btautilde{{\widetilde\btau}}
\def\bupsilontilde{{\widetilde\bupsilon}}
\def\bphitilde{{\widetilde\bphi}}
\def\bchitilde{{\widetilde\bchi}}
\def\bpsitilde{{\widetilde\bpsi}}
\def\bomegatilde{{\widetilde\bomega}}
\def\bAlphatilde{{\widetilde\bAlpha}}
\def\bBetatilde{{\widetilde\bBeta}}
\def\bGammatilde{{\widetilde\bGamma}}
\def\bDeltatilde{{\widetilde\bDelta}}
\def\bEpsilontilde{{\widetilde\bEpsilon}}
\def\bZetatilde{{\widetilde\bZeta}}
\def\bEtatilde{{\widetilde\bEta}}
\def\bThetatilde{{\widetilde\bTheta}}
\def\bIotatilde{{\widetilde\bIota}}
\def\bKappatilde{{\widetilde\bKappa}}
\def\bLambdatilde{{\widetilde\bLambda}}
\def\bMutilde{{\widetilde\bMu}}
\def\bNutilde{{\widetilde\bNu}}
\def\bXitilde{{\widetilde\bXi}}
\def\bOmicrontilde{{\widetilde\bOmicron}}
\def\bPitilde{{\widetilde\bPi}}
\def\bRhotilde{{\widetilde\bRho}}
\def\bSigmatilde{{\widetilde\bSigma}}
\def\bTautilde{{\widetilde\bTau}}
\def\bUpsilontilde{{\widetilde\bUpsilon}}
\def\bPhitilde{{\widetilde\bPhi}}
\def\bChitilde{{\widetilde\bChi}}
\def\bPsitilde{{\widetilde\bPsi}}
\def\bOmegatilde{{\widetilde\bOmega}}
%
%
%
%
%
\def\abar{\bar{ a}}
\def\bbar{\bar{ b}}
\def\cbar{\bar{ c}}
\def\dbar{\bar{ d}}
\def\ebar{\bar{ e}}
\def\fbar{\bar{ f}}
\def\gbar{\bar{ g}}
\def\hbar{\bar{ h}}
\def\ibar{\bar{ i}}
\def\jbar{\bar{ j}}
\def\kbar{\bar{ k}}
\def\lbar{\bar{ l}}
\def\mbar{\bar{ m}}
\def\nbar{\bar{ n}}
\def\obar{\bar{ o}}
\def\pbar{\bar{ p}}
\def\qbar{\bar{ q}}
\def\rbar{\bar{ r}}
\def\sbar{\bar{ s}}
\def\tbar{\bar{ t}}
\def\ubar{\bar{ u}}
\def\vbar{\bar{ v}}
\def\wbar{\bar{ w}}
\def\xbar{\bar{ x}}
\def\ybar{\bar{ y}}
\def\zbar{\bar{ z}}
\def\Abar{\bar{ A}}
\def\Bbar{\bar{ B}}
\def\Cbar{\bar{ C}}
\def\Dbar{\bar{ D}}
\def\Ebar{\bar{ E}}
\def\Fbar{\bar{ F}}
\def\Gbar{\bar{ G}}
\def\Hbar{\bar{ H}}
\def\Ibar{\bar{ I}}
\def\Jbar{\bar{ J}}
\def\Kbar{\bar{ K}}
\def\Lbar{\bar{ L}}
\def\Mbar{\bar{ M}}
\def\Nbar{\bar{ N}}
\def\Obar{\bar{ O}}
\def\Pbar{\bar{ P}}
\def\Qbar{\bar{ Q}}
\def\Rbar{\bar{ R}}
\def\Sbar{\bar{ S}}
\def\Tbar{\bar{ T}}
\def\Ubar{\bar{ U}}
\def\Vbar{\bar{ V}}
\def\Wbar{\bar{ W}}
\def\Xbar{\bar{ X}}
\def\Ybar{\bar{ Y}}
\def\Zbar{\bar{ Z}}
%
%
%
%
%
\def\babar{\overline{ \ba}}
\def\bbbar{\overline{ \bb}}
\def\bcbar{\overline{ \bc}}
\def\bdbar{\overline{ \bd}}
\def\bebar{\overline{ \be}}
\def\bfbar{\overline{ \bf}}
\def\bgbar{\overline{ \bg}}
\def\bhbar{\overline{ \bh}}
\def\bibar{\overline{ \bi}}
\def\bjbar{\overline{ \bj}}
\def\bkbar{\overline{ \bk}}
\def\blbar{\overline{ \bl}}
\def\bmbar{\overline{ \bm}}
\def\bnbar{\overline{ \bn}}
\def\bobar{\overline{ \bo}}
\def\bpbar{\overline{ \bp}}
\def\bqbar{\overline{ \bq}}
\def\brbar{\overline{ \br}}
\def\bsbar{\overline{ \bs}}
\def\btbar{\overline{ \bt}}
\def\bubar{\overline{ \bu}}
\def\bvbar{\overline{ \bv}}
\def\bwbar{\overline{ \bw}}
\def\bxbar{\overline{ \bx}}
\def\bybar{\overline{ \by}}
\def\bzbar{\overline{ \bz}}
\def\bAbar{\overline{ \bA}}
\def\bBbar{\overline{ \bB}}
\def\bCbar{\overline{ \bC}}
\def\bDbar{\overline{ \bD}}
\def\bEbar{\overline{ \bE}}
\def\bFbar{\overline{ \bF}}
\def\bGbar{\overline{ \bG}}
\def\bHbar{\overline{ \bH}}
\def\bIbar{\overline{ \bI}}
\def\bJbar{\overline{ \bJ}}
\def\bKbar{\overline{ \bK}}
\def\bLbar{\overline{ \bL}}
\def\bMbar{\overline{ \bM}}
\def\bNbar{\overline{ \bN}}
\def\bObar{\overline{ \bO}}
\def\bPbar{\overline{ \bP}}
\def\bQbar{\overline{ \bQ}}
\def\bRbar{\overline{ \bR}}
\def\bSbar{\overline{ \bS}}
\def\bTbar{\overline{ \bT}}
\def\bUbar{\overline{ \bU}}
\def\bVbar{\overline{ \bV}}
\def\bWbar{\overline{ \bW}}
\def\bXbar{\overline{ \bX}}
\def\bYbar{\overline{ \bY}}
\def\bZbar{\overline{ \bZ}}
%
%

%
%
%
\def\asc{{\cal a}}
\def\bsc{{\cal b}}
\def\csc{{\cal c}}
\def\dsc{{\cal d}}
\def\esc{{\cal e}}
\def\dsc{{\cal f}}
\def\gsc{{\cal g}}
\def\hsc{{\cal h}}
\def\isc{{\cal i}}
\def\jsc{{\cal j}}
\def\ksc{{\cal k}}
\def\lsc{{\cal l}}
\def\msc{{\cal m}}
\def\nsc{{\cal n}}
\def\osc{{\cal o}}
\def\psc{{\cal p}}
\def\qsc{{\cal q}}
\def\rsc{{\cal r}}
\def\ssc{{\cal s}}
\def\tsc{{\cal t}}
\def\usc{{\cal u}}
\def\vsc{{\cal v}}
\def\wsc{{\cal w}}
\def\xsc{{\cal x}}
\def\ysc{{\cal y}}
\def\zsc{{\cal z}}
\def\Asc{{\cal A}}
\def\Bsc{{\cal B}}
\def\Csc{{\cal C}}
\def\Dsc{{\cal D}}
\def\Esc{{\cal E}}
\def\Fsc{{\cal F}}
\def\Gsc{{\cal G}}
\def\Hsc{{\cal H}}
\def\Isc{{\cal I}}
\def\Jsc{{\cal J}}
\def\Ksc{{\cal K}}
\def\Lsc{{\cal L}}
\def\Msc{{\cal M}}
\def\Nsc{{\cal N}}
\def\Osc{{\cal O}}
\def\Psc{{\cal P}}
\def\Qsc{{\cal Q}}
\def\Rsc{{\cal R}}
\def\Ssc{{\cal S}}
\def\Tsc{{\cal S}}
\def\Usc{{\cal U}}
\def\Vsc{{\cal V}}
\def\Wsc{{\cal W}}
\def\Xsc{{\cal X}}
\def\Ysc{{\cal Y}}
\def\Zsc{{\cal Z}}
\def\Aschat{\widehat{{\cal A}}}
\def\Bschat{\widehat{{\cal B}}}
\def\Cschat{\widehat{{\cal C}}}
\def\Dschat{\widehat{{\cal D}}}
\def\Eschat{\widehat{{\cal E}}}
\def\Fschat{\widehat{{\cal F}}}
\def\Gschat{\widehat{{\cal G}}}
\def\Hschat{\widehat{{\cal H}}}
\def\Ischat{\widehat{{\cal I}}}
\def\Jschat{\widehat{{\cal J}}}
\def\Kschat{\widehat{{\cal K}}}
\def\Lschat{\widehat{{\cal L}}}
\def\Mschat{\widehat{{\cal M}}}
\def\Nschat{\widehat{{\cal N}}}
\def\Oschat{\widehat{{\cal O}}}
\def\Pschat{\widehat{{\cal P}}}
\def\Qschat{\widehat{{\cal Q}}}
\def\Rschat{\widehat{{\cal R}}}
\def\Sschat{\widehat{{\cal S}}}
\def\Tschat{\widehat{{\cal T}}}
\def\Uschat{\widehat{{\cal U}}}
\def\Vschat{\widehat{{\cal V}}}
\def\Wschat{\widehat{{\cal W}}}
\def\Xschat{\widehat{{\cal X}}}
\def\Yschat{\widehat{{\cal Y}}}
\def\Zschat{\widehat{{\cal Z}}}
\def\Asctilde{\widetilde{{\cal A}}}
\def\Bsctilde{\widetilde{{\cal B}}}
\def\Csctilde{\widetilde{{\cal C}}}
\def\Dsctilde{\widetilde{{\cal D}}}
\def\Esctilde{\widetilde{{\cal E}}}
\def\Fsctilde{\widetilde{{\cal F}}}
\def\Gsctilde{\widetilde{{\cal G}}}
\def\Hsctilde{\widetilde{{\cal H}}}
\def\Isctilde{\widetilde{{\cal I}}}
\def\Jsctilde{\widetilde{{\cal J}}}
\def\Ksctilde{\widetilde{{\cal K}}}
\def\Lsctilde{\widetilde{{\cal L}}}
\def\Msctilde{\widetilde{{\cal M}}}
\def\Nsctilde{\widetilde{{\cal N}}}
\def\Osctilde{\widetilde{{\cal O}}}
\def\Psctilde{\widetilde{{\cal P}}}
\def\Qsctilde{\widetilde{{\cal Q}}}
\def\Rsctilde{\widetilde{{\cal R}}}
\def\Ssctilde{\widetilde{{\cal S}}}
\def\Tsctilde{\widetilde{{\cal T}}}
\def\Usctilde{\widetilde{{\cal U}}}
\def\Vsctilde{\widetilde{{\cal V}}}
\def\Wsctilde{\widetilde{{\cal W}}}
\def\Xsctilde{\widetilde{{\cal X}}}
\def\Ysctilde{\widetilde{{\cal Y}}}
\def\Zsctilde{\widetilde{{\cal Z}}}
\def\bAsc{\mathbf{\cal A}}
\def\bBsc{\mathbf{\cal B}}
\def\bCsc{\mathbf{\cal C}}
\def\bDsc{\mathbf{\cal D}}
\def\bEsc{\mathbf{\cal E}}
\def\bFsc{\mathbf{\cal F}}
\def\bGsc{\mathbf{\cal G}}
\def\bHsc{\mathbf{\cal H}}
\def\bIsc{\mathbf{\cal I}}
\def\bJsc{\mathbf{\cal J}}
\def\bKsc{\mathbf{\cal K}}
\def\bLsc{\mathbf{\cal L}}
\def\bMsc{\mathbf{\cal M}}
\def\bNsc{\mathbf{\cal N}}
\def\bOsc{\mathbf{\cal O}}
\def\bPsc{\mathbf{\cal P}}
\def\bQsc{\mathbf{\cal Q}}
\def\bRsc{\mathbf{\cal R}}
\def\bSsc{\mathbf{\cal S}}
\def\bTsc{\mathbf{\cal T}}
\def\bUsc{\mathbf{\cal U}}
\def\bVsc{\mathbf{\cal V}}
\def\bWsc{\mathbf{\cal W}}
\def\bXsc{\mathbf{\cal X}}
\def\bYsc{\mathbf{\cal Y}}
\def\bZsc{\mathbf{\cal Z}}
\def\bAschat{\widehat{\mathbf{\cal A}}}
\def\bBschat{\widehat{\mathbf{\cal B}}}
\def\bCschat{\widehat{\mathbf{\cal C}}}
\def\bDschat{\widehat{\mathbf{\cal D}}}
\def\bEschat{\widehat{\mathbf{\cal E}}}
\def\bFschat{\widehat{\mathbf{\cal F}}}
\def\bGschat{\widehat{\mathbf{\cal G}}}
\def\bHschat{\widehat{\mathbf{\cal H}}}
\def\bIschat{\widehat{\mathbf{\cal I}}}
\def\bJschat{\widehat{\mathbf{\cal J}}}
\def\bKschat{\widehat{\mathbf{\cal K}}}
\def\bLschat{\widehat{\mathbf{\cal L}}}
\def\bMschat{\widehat{\mathbf{\cal M}}}
\def\bNschat{\widehat{\mathbf{\cal N}}}
\def\bOschat{\widehat{\mathbf{\cal O}}}
\def\bPschat{\widehat{\mathbf{\cal P}}}
\def\bQschat{\widehat{\mathbf{\cal Q}}}
\def\bRschat{\widehat{\mathbf{\cal R}}}
\def\bSschat{\widehat{\mathbf{\cal S}}}
\def\bTschat{\widehat{\mathbf{\cal T}}}
\def\bUschat{\widehat{\mathbf{\cal U}}}
\def\bVschat{\widehat{\mathbf{\cal V}}}
\def\bWschat{\widehat{\mathbf{\cal W}}}
\def\bXschat{\widehat{\mathbf{\cal X}}}
\def\bYschat{\widehat{\mathbf{\cal Y}}}
\def\bZschat{\widehat{\mathbf{\cal Z}}}
\def\afrak{\mathfrak{a}}
\def\bfrak{\mathfrak{b}}
\def\cfrak{\mathfrak{c}}
\def\dfrak{\mathfrak{d}}
\def\efrak{\mathfrak{e}}
\def\ffrak{\mathfrak{f}}
\def\gfrak{\mathfrak{g}}
\def\hfrak{\mathfrak{h}}
\def\ifrak{\mathfrak{i}}
\def\jfrak{\mathfrak{j}}
\def\kfrak{\mathfrak{k}}
\def\lfrak{\mathfrak{l}}
\def\mfrak{\mathfrak{m}}
\def\nfrak{\mathfrak{n}}
\def\ofrak{\mathfrak{o}}
\def\pfrak{\mathfrak{p}}
\def\qfrak{\mathfrak{q}}
\def\rfrak{\mathfrak{r}}
\def\sfrak{\mathfrak{s}}
\def\tfrak{\mathfrak{t}}
\def\ufrak{\mathfrak{u}}
\def\vfrak{\mathfrak{v}}
\def\wfrak{\mathfrak{w}}
\def\xfrak{\mathfrak{x}}
\def\yfrak{\mathfrak{y}}
\def\zfrak{\mathfrak{z}}
\def\Afrak{\mathfrak{ A}}
\def\Bfrak{\mathfrak{ B}}
\def\Cfrak{\mathfrak{ C}}
\def\Dfrak{\mathfrak{ D}}
\def\Efrak{\mathfrak{ E}}
\def\Ffrak{\mathfrak{ F}}
\def\Gfrak{\mathfrak{ G}}
\def\Hfrak{\mathfrak{ H}}
\def\Ifrak{\mathfrak{ I}}
\def\Jfrak{\mathfrak{ J}}
\def\Kfrak{\mathfrak{ K}}
\def\Lfrak{\mathfrak{ L}}
\def\Mfrak{\mathfrak{ M}}
\def\Nfrak{\mathfrak{ N}}
\def\Ofrak{\mathfrak{ O}}
\def\Pfrak{\mathfrak{ P}}
\def\Qfrak{\mathfrak{ Q}}
\def\Rfrak{\mathfrak{ R}}
\def\Sfrak{\mathfrak{ S}}
\def\Tfrak{\mathfrak{ T}}
\def\Ufrak{\mathfrak{ U}}
\def\Vfrak{\mathfrak{ V}}
\def\Wfrak{\mathfrak{ W}}
\def\Xfrak{\mathfrak{ X}}
\def\Yfrak{\mathfrak{ Y}}
\def\Zfrak{\mathfrak{ Z}}

\def\bAfrak{\mathbf{\mathfrak{A}}}
\def\bBfrak{\mathbf{\mathfrak{B}}}
\def\bCfrak{\mathbf{\mathfrak{C}}}
\def\bDfrak{\mathbf{\mathfrak{D}}}
\def\bEfrak{\mathbf{\mathfrak{E}}}
\def\bFfrak{\mathbf{\mathfrak{F}}}
\def\bGfrak{\mathbf{\mathfrak{G}}}
\def\bHfrak{\mathbf{\mathfrak{H}}}
\def\bIfrak{\mathbf{\mathfrak{I}}}
\def\bJfrak{\mathbf{\mathfrak{J}}}
\def\bKfrak{\mathbf{\mathfrak{K}}}
\def\bLfrak{\mathbf{\mathfrak{L}}}
\def\bMfrak{\mathbf{\mathfrak{M}}}
\def\bNfrak{\mathbf{\mathfrak{N}}}
\def\bOfrak{\mathbf{\mathfrak{O}}}
\def\bPfrak{\mathbf{\mathfrak{P}}}
\def\bQfrak{\mathbf{\mathfrak{Q}}}
\def\bRfrak{\mathbf{\mathfrak{R}}}
\def\bSfrak{\mathbf{\mathfrak{S}}}
\def\bTfrak{\mathbf{\mathfrak{T}}}
\def\bUfrak{\mathbf{\mathfrak{U}}}
\def\bVfrak{\mathbf{\mathfrak{V}}}
\def\bWfrak{\mathbf{\mathfrak{W}}}
\def\bXfrak{\mathbf{\mathfrak{X}}}
\def\bYfrak{\mathbf{\mathfrak{Y}}}
\def\bZfrak{\mathbf{\mathfrak{Z}}}

\def\bAfrakhat{\mathbf{\widehat{\mathfrak{A}}}}
\def\bBfrakhat{\mathbf{\widehat{\mathfrak{B}}}}
\def\bCfrakhat{\mathbf{\widehat{\mathfrak{C}}}}
\def\bDfrakhat{\mathbf{\widehat{\mathfrak{D}}}}
\def\bEfrakhat{\mathbf{\widehat{\mathfrak{E}}}}
\def\bFfrakhat{\mathbf{\widehat{\mathfrak{F}}}}
\def\bGfrakhat{\mathbf{\widehat{\mathfrak{G}}}}
\def\bHfrakhat{\mathbf{\widehat{\mathfrak{H}}}}
\def\bIfrakhat{\mathbf{\widehat{\mathfrak{I}}}}
\def\bJfrakhat{\mathbf{\widehat{\mathfrak{J}}}}
\def\bKfrakhat{\mathbf{\widehat{\mathfrak{K}}}}
\def\bLfrakhat{\mathbf{\widehat{\mathfrak{L}}}}
\def\bMfrakhat{\mathbf{\widehat{\mathfrak{M}}}}
\def\bNfrakhat{\mathbf{\widehat{\mathfrak{N}}}}
\def\bOfrakhat{\mathbf{\widehat{\mathfrak{O}}}}
\def\bPfrakhat{\mathbf{\widehat{\mathfrak{P}}}}
\def\bQfrakhat{\mathbf{\widehat{\mathfrak{Q}}}}
\def\bRfrakhat{\mathbf{\widehat{\mathfrak{R}}}}
\def\bSfrakhat{\mathbf{\widehat{\mathfrak{S}}}}
\def\bTfrakhat{\mathbf{\widehat{\mathfrak{T}}}}
\def\bUfrakhat{\mathbf{\widehat{\mathfrak{U}}}}
\def\bVfrakhat{\mathbf{\widehat{\mathfrak{V}}}}
\def\bWfrakhat{\mathbf{\widehat{\mathfrak{W}}}}
\def\bXfrakhat{\mathbf{\widehat{\mathfrak{X}}}}
\def\bYfrakhat{\mathbf{\widehat{\mathfrak{Y}}}}
\def\bZfrakhat{\mathbf{\widehat{\mathfrak{Z}}}}
%
%
%
%
\def\etal{{\em et al.}}
%
%
%
%
%
\def\cumsum{\mbox{cumsum}}
\def\real{{\mathbb R}}
\def\intinfinf{\int_{-\infty}^{\infty}}
\def\intzinf{\int_{0}^{\infty}}
\def\intzt{\int_0^t}
\def\transpose{{\sf \scriptscriptstyle{T}}}
\def\smhalf{{\textstyle{1\over2}}}
\def\third{{\textstyle{1\over3}}}
\def\twothirds{{\textstyle{2\over3}}}
\def\bell{\bmath{\ell}}
\def\half{\frac{1}{2}}
\def\ninv{n^{-1}}
\def\nhalf{n^{\half}}
\def\mhalf{m^{\half}}
\def\nnhalf{n^{-\half}}
\def\mnhalf{m^{-\half}}
\def\MN{\mbox{MN}}
\def\N{\mbox{N}}
\def\E{\mbox{E}}
\def\pr{P}
\def\var{\mbox{var}}
\def\limn{\lim_{n\to \infty} }
\def\intt{\int_{\tau_a}^{\tau_b}}
\def\sumin{\sum_{i=1}^n}
\def\sumjn{\sum_{j=1}^n}
\def\SUMin{{\displaystyle \sum_{i=1}^n}}
\def\SUMjn{{\displaystyle \sum_{j=1}^n}}
\def\myendthm{\begin{flushright} $\diamond $ \end{flushright}}
\def\convd{\overset{\Dsc}{\longrightarrow}}
\def\convp{\overset{\Psc}{\longrightarrow}}
\def\convas{\overset{a.s.}{\longrightarrow}}
\def\hn{\mbox{H}_0}
\def\ha{\mbox{H}_1}

%
%
%
%
%
\def\trans{^{\transpose}}
\def\inv{^{-1}}
\def\twobyone#1#2{\left[
\begin{array}
{c}
#1\\
#2\\
\end{array}
\right]}
%
%
%
%
%
\def\argmindum{\mathop{\mbox{argmin}}}
\def\argmin#1{\argmindum_{#1}}
\def\argmaxdum{\mathop{\mbox{argmax}}}
\def\argmax#1{\argmaxdum_{#1}}
\def\blockdiag{\mbox{blockdiag}}
\def\corr{\mbox{corr}}
\def\cov{\mbox{cov}}
\def\diag{\mbox{diag}}
\def\dffit{df_{{\rm fit}}}
\def\dfres{df_{{\rm res}}}
\def\dfyhat{df_{\yhat}}
\def\diag{\mbox{diag}}
\def\diagonal{\mbox{diagonal}}
\def\logit{\mbox{logit}}
\def\stdev{\mbox{st.\,dev.}}
\def\stdevhat{{\widehat{\mbox{st.dev}}}}
\def\tr{\mbox{tr}}
\def\trigamma{\mbox{trigamma}}
\def\var{\mbox{var}}
\def\vecof{\mbox{vec}}
\def\AIC{\mbox{AIC}}
\def\AMISE{\mbox{AMISE}}
\def\Corr{\mbox{Corr}}
\def\Cov{\mbox{Cov}}
\def\CV{\mbox{CV}}
\def\GCV{\mbox{GCV}}
\def\LR{\mbox{LR}}
\def\MISE{\mbox{MISE}}
\def\MSSE{\mbox{MSSE}}
\def\ML{\mbox{ML}}
\def\REML{\mbox{REML}}
\def\RMSE{{\rm RMSE}}
\def\RSS{\mbox{RSS}}
\def\Var{\mbox{Var}}
%
%
%
%
\def\bib{\vskip12pt\par\noindent\hangindent=1 true cm\hangafter=1}
\def\jump{\vskip3mm\noindent}
\def\mybox#1{\vskip1mm \begin{center}
        \hspace{.0\textwidth}\vbox{\hrule\hbox{\vrule\kern6pt
\parbox{.9\textwidth}{\kern6pt#1\vskip6pt}\kern6pt\vrule}\hrule}
        \end{center} \vskip-5mm}
\def\lboxit#1{\vbox{\hrule\hbox{\vrule\kern6pt
      \vbox{\kern6pt#1\vskip6pt}\kern6pt\vrule}\hrule}}
\def\boxit#1{\begin{center}\fbox{#1}\end{center}}
\def\thickboxit#1{\vbox{{\hrule height 1mm}\hbox{{\vrule width 1mm}\kern6pt
          \vbox{\kern6pt#1\kern6pt}\kern6pt{\vrule width 1mm}}
               {\hrule height 1mm}}}
\def\instep{\vskip12pt\par\hangindent=30 true mm\hangafter=1}
\def\uWand{\underline{Wand}}
\def\remtask#1#2{\mmnote{\thickboxit
                 {\bf #1\ \theremtask}}\refstepcounter{remtask}}
%
%
%

%
%
\def\aism{{\it Ann. Inst. Statist. Math.}\ }
\def\ajs{{\it Austral. J. Statist.}\ }
\def\ANNSTAT{{\it The Annals of Statistics}\ }
\def\annmath{{\it Ann. Math. Statist.}\ }
\def\applstat{{\it Appl. Statist.}\ }
\def\BIOMETRICS{{\it Biometrics}\ }
\def\cjs{{\it Canad. J. Statist.}\ }
\def\csda{{\it Comp. Statist. Data Anal.}\ }
\def\cstm{{\it Comm. Statist. Theory Meth.}\ }
\def\ieeetit{{\it IEEE Trans. Inf. Theory}\ }
\def\isr{{\it Internat. Statist. Rev.}\ }
\def\JASA{{\it Journal of the American Statistical Association}\ }
\def\JCGS{{\it Journal of Computational and Graphical Statistics}\ }
\def\jscs{{\it J. Statist. Comput. Simulation}\ }
\def\jma{{\it J. Multivariate Anal.}\ }
\def\jns{{\it J. Nonparametric Statist.}\ }
\def\JRSSA{{\it Journal of the Royal Statistics Society, Series A}\ }
\def\JRSSB{{\it Journal of the Royal Statistics Society, Series B}\ }
\def\JRSSC{{\it Journal of the Royal Statistics Society, Series C}\ }
\def\jspi{{\it J. Statist. Planning Inference}\ }
\def\ptrf{{\it Probab. Theory Rel. Fields}\ }
\def\sankhyaa{{\it Sankhy$\bar{{\it a}}$} Ser. A\ }
\def\sjs{{\it Scand. J. Statist.}\ }
\def\spl{{\it Statist. Probab. Lett.}\ }
\def\statsci{{\it Statist. Sci.}\ }
\def\techno{{\it Technometrics}\ }
\def\tpa{{\it Theory Probab. Appl.}\ }
\def\zw{{\it Z. Wahr. ver. Geb.}\ }
%
%
%
%
\def\Brent{{\bf BRENT:}\ }
\def\David{{\bf DAVID:}\ }
\def\Erin{{\bf ERIN:}}
\def\Gerda{{\bf GERDA:}\ }
\def\Joel{{\bf JOEL:}\ }
\def\Marc{{\bf MARC:}\ }
\def\Matt{{\bf MATT:}\ }
\def\Tianxi{{\bf TIANXI:}\ }
%
%
%
%
\def\bZE{\bZ_{\scriptscriptstyle E}}
\def\bZT{\bZ_{\scriptscriptstyle T}}
\def\bbE{\bb_{\scriptscriptstyle E}}
\def\bbT{\bb_{\scriptscriptstyle T}}
\def\bbhatT{\bbhat_{\scriptscriptstyle T}}
\def\fX{f_{\scriptscriptstyle X}}
\def\sigeps{\sigma_{\varepsilon}}
\def\bVtheta{\bV_{\smbtheta}}
\def\bVthetainv{\bVtheta^{-1}}
\def\bKsc{\boldsymbol{\Ksc}}
\def\bxbar{\bar{\bx}}
\def\bPL{b^{\scriptscriptstyle{\rm PL}}}
\def\bVA{b^{\scriptscriptstyle{\rm VA}}}
\def\zPL{z^{\scriptscriptstyle{\rm PL}}}
\def\zVA{z^{\scriptscriptstyle{\rm VA}}}
\def\bYmis{\bY_{\scriptscriptstyle{\rm mis}}}
\def\bYmishat{{\widehat{\bYmis}}}
\def\bYmisone{\bY_{\scriptscriptstyle{\rm mis,1}}}
\def\bYmistwo{\bY_{\scriptscriptstyle{\rm mis,2}}}
\def\bYobs{\bY_{\scriptscriptstyle{\rm obs}}}
\def\bdobs{\bd_{\scriptscriptstyle{\rm obs}}}
\def\bdmis{\bd_{\scriptscriptstyle{\rm mis}}}
%
%
%
%
\def\bfDelta{{\mbox{\boldmath$\Delta$}}}
\def\bfkappa{{\mbox{\boldmath$\kappa$}}}
\def\bfgamma{{\mbox{\boldmath$\gamma$}}}
\def\bftheta{{\mbox{\boldmath$\theta$}}}
\def\bfmu{{\mbox{\boldmath$\mu$}}}
\def\bfdelta{{\mbox{\boldmath$\delta$}}}
\def\bfeps{{\mbox{\boldmath$\varepsilon$}}}
\def\bfnu{{\mbox{\boldmath$\nu$}}}
\def\bfzeta{{\mbox{\boldmath$\zeta$}}}
\def\bfchi{{\mbox{\boldmath$\chi$}}}
\def\bbX{\mathbb{X}}
\def\bbV{\mathbb{V}} 
\def\bbA{\mathbb{A}}
\def\bbB{\mathbb{B}}
\def\bbK{\mathbb{K}}
\def\bbP{\mathbb{P}}
\def\bbD{\mathbb{D}}

\def\Abb{\mathbb{A}}
\def\Bbb{\mathbb{B}}
\def\Cbb{\mathbb{C}}
\def\Dbb{\mathbb{D}}
\def\Ebb{\mathbb{E}}
\def\Fbb{\mathbb{F}}
\def\Gbb{\mathbb{G}}
\def\Hbb{\mathbb{H}}
\def\Ibb{\mathbb{I}}
\def\Jbb{\mathbb{J}}
\def\Kbb{\mathbb{K}}
\def\Lbb{\mathbb{L}}
\def\Mbb{\mathbb{M}}
\def\Nbb{\mathbb{N}}
\def\Mbb{\mathbb{M}}
\def\Nbb{\mathbb{N}}
\def\Obb{\mathbb{O}}
\def\Pbb{\mathbb{P}}
\def\Qbb{\mathbb{Q}}
\def\Rbb{\mathbb{R}}
\def\Sbb{\mathbb{S}}
\def\Tbb{\mathbb{T}}
\def\Ubb{\mathbb{U}}
\def\Vbb{\mathbb{V}}
\def\Wbb{\mathbb{W}}
\def\Xbb{\mathbb{X}}
\def\Ybb{\mathbb{Y}}
\def\Zbb{\mathbb{Z}}

\def\Abbtilde{\widetilde{\mathbb{A}}}
\def\Bbbtilde{\widetilde{\mathbb{B}}}
\def\Cbbtilde{\widetilde{\mathbb{C}}}
\def\Dbbtilde{\widetilde{\mathbb{D}}}
\def\Ebbtilde{\widetilde{\mathbb{E}}}
\def\Fbbtilde{\widetilde{\mathbb{F}}}
\def\Gbbtilde{\widetilde{\mathbb{G}}}
\def\Hbbtilde{\widetilde{\mathbb{H}}}
\def\Ibbtilde{\widetilde{\mathbb{I}}}
\def\Jbbtilde{\widetilde{\mathbb{J}}}
\def\Kbbtilde{\widetilde{\mathbb{K}}}
\def\Lbbtilde{\widetilde{\mathbb{L}}}
\def\Mbbtilde{\widetilde{\mathbb{M}}}
\def\Nbbtilde{\widetilde{\mathbb{N}}}
\def\Mbbtilde{\widetilde{\mathbb{M}}}
\def\Nbbtilde{\widetilde{\mathbb{N}}}
\def\Obbtilde{\widetilde{\mathbb{O}}}
\def\Pbbtilde{\widetilde{\mathbb{P}}}
\def\Qbbtilde{\widetilde{\mathbb{Q}}}
\def\Rbbtilde{\widetilde{\mathbb{R}}}
\def\Sbbtilde{\widetilde{\mathbb{S}}}
\def\Tbbtilde{\widetilde{\mathbb{T}}}
\def\Ubbtilde{\widetilde{\mathbb{U}}}
\def\Vbbtilde{\widetilde{\mathbb{V}}}
\def\Wbbtilde{\widetilde{\mathbb{W}}}
\def\Xbbtilde{\widetilde{\mathbb{X}}}
\def\Ybbtilde{\widetilde{\mathbb{Y}}}
\def\Zbbtilde{\widetilde{\mathbb{Z}}}

%
%
%
%
\def\miss{\mbox{{\tiny miss}}}
\def\obs{\scriptsize{\mbox{obs}}}

%
%
%
%
\def\bmath#1{\mbox{\boldmath$#1$}}
\def\fat#1{\hbox{\rlap{$#1$}\kern0.25pt\rlap{$#1$}\kern0.25pt$#1$}}
\def\wh{\widehat}
\def\flambda{\fat{\lambda}}
\def\beps{\bmath{\varepsilon}}
\def\bSlambda{\bS_{\lambda}}
\def\ErrorSS{\mbox{RSS}}
\def\bsqbar{\bar{{b^2}}}
\def\bcubar{\bar{{b^3}}}
\def\plargest{p_{\rm \,largest}}
\def\summheading#1{\subsection*{#1}\hskip3mm}
\def\summbreak{\vskip3mm\par}
\def\df{df}
\def\adf{adf}
\def\dffit{df_{{\rm fit}}}
\def\dfres{df_{{\rm res}}}
\def\dfyhat{df_{\yhat}}
\def\sigb{\sigma_b}
\def\sigu{\sigma_u}
\def\sigepshat{{\widehat\sigma}_{\varepsilon}}
\def\siguhat{{\widehat\sigma}_u}
\def\sigepshat{{\widehat\sigma}_{\varepsilon}}
\def\sigbhat{{\widehat\sigma}_b}
\def\sighat{{\widehat\sigma}}
\def\sigsqb{\sigma^2_b}
\def\sigsqeps{\sigma^2_{\varepsilon}}
\def\sigsqepszerohat{{\widehat\sigma}^2_{\varepsilon,0}}
\def\sigsqepshat{{\widehat\sigma}^2_{\varepsilon}}
\def\sigsqbhat{{\widehat\sigma}^2_b}
\def\dfnumer{{\rm df(II}|{\rm I)}}
\def\mhatlam{{\widehat m}_{\lambda}}
\def\calD{\Dsc}
\def\Aeps{A_{\epsilon}}
\def\Beps{B_{\epsilon}}
\def\Ab{A_b}
\def\Bb{B_b}
\def\bXtmain{\tilde{\bX}_r}
\def\main{\mbox{\tt main}}
\def\argminbetab{\argmin{\bbeta,\bb}}
\def\calB{\Bsc}
\def\respvar{\mbox{\tt log(amt)}}

\def\Abb{\mathbb{A}}
\def\Zbb{\mathbb{Z}}
\def\Wbb{\mathbb{W}}
\def\Wbbhat{\widehat{\mathbb{W}}}
\def\Kbbtilde{\widetilde{\mathbb{K}}}
\def\Pbbtilde{\widetilde{\mathbb{P}}}
\def\Dbbtilde{\widetilde{\mathbb{D}}}
\def\Bbbtilde{\widetilde{\mathbb{B}}}

\def\Abbhat{\widehat{\mathbb{A}}}

\def\ellhat{\widehat{\ell}}
\def\pn{\phantom{-}}
\def\pp{\phantom{1}}

\newtheorem{definition}{Definition}
\newtheorem{lemma}{Lemma}
\newtheorem{lemmaA}{Lemma}
\newtheorem{prop}{Proposition}
\newtheorem{assume}{Assumption}
\newtheorem{cond}{Condition}
\newtheorem{coro}{Corollary}
\newtheorem{theorem}{Theorem}
\newtheorem{remark}{Remark}
\newtheorem{claim}{Claim}

\def\PP{\stackrel{P}{\rightarrow}}
\def\DD{\Rightarrow}
%
%

\ifjournal
\volume{}
\artmonth{}
\else
\maketitle
\date{}

\fi

\begin{abstract}

With the increasing availability of electronic health records (EHR) linked with biobank data for translational research, a critical step in realizing its potential is to accurately classify phenotypes for patients. Existing approaches to achieve this goal are based on error-prone EHR surrogate outcomes, assisted and validated by a small set of labels obtained via medical chart review, which may also be subject to misclassification. Ignoring the noise in these outcomes can induce severe estimation and validation bias to both EHR phenotyping and risking modeling with biomarkers collected in the biobank. To overcome this challenge, we propose a novel unsupervised and semiparametric approach to jointly model multiple noisy EHR outcomes with their linked biobank features. Our approach primarily aims at disease risk modeling with the baseline biomarkers, and is also able to produce a predictive EHR phenotyping model and validate its performance without observations of the true disease outcome. It consists of composite and nonparametric regression steps free of any parametric model specification, followed by a parametric projection step to reduce the uncertainty and improve the estimation efficiency. We show that our method is robust to violations of the parametric assumptions while attaining the desirable root-$n$ convergence rates on risk modeling. Our developed method outperforms existing methods in extensive simulation studies, as well as a real-world application in phenotyping and genetic risk modeling of type II diabetes.


\end{abstract}

\noindent{\bf Keywords}: EHR linked biobank data; Surrogates; Measurement errors; Biomarker; Model misspecification; Under-smoothing.


\section{Introduction}

\subsection{Background}
With the increasing adoption of electronic health record (EHR) systems in the United States,  EHR data are increasingly accessible for research. Linking EHR data with biorepository, powerful phenome-genome studies can be performed with such large scale data for discovery and translational research  \citep{kohane2011using,denny2013systematic,wells2019accelerating}. To fully realize the potential of EHR data, a critical step involves accurately and efficiently classifying phenotype status for individual patients to enable association studies and risk modeling. Although simple rule-based classification algorithms leveraging domain knowledge remain useful, they have varying degree of accuracy and portability \citep{zhang2019high}. Conversely, data-driven machine learning based classification algorithms have been advocated as a useful alternative with higher accuracy and portability \citep{shivade2014review,liao2015development,banda2018advances}. Typically, these algorithms undergo training and/or validation with gold standard labels curated via medical chart review. Subsequently, the predicted phenotypes for all patients in the cohort serve as the observed outcomes for downstream association studies \citep[e.g]{liao2013associations,map2019}. 

Historically, most existing phenotyping algorithms have relied on supervised methods, which suffer from scalablility issue due to labor-intensive nature of manually reviewing charts to obtain gold standard labels for the phenotype of interest. In recent years, several unsupervised methods leveraging unlabeled data using surrogate features as noisy labels \citep{yu2017enabling,banda2017electronic,map2019} were proposed as promising alternatives. However, these methods can lead to poor accuracy when the surrogate features have limited accuracy and do not provide reliable estimate of classification performance of the trained models.

\subsection{Problem setup}\label{sec:model}

Let $Y$ denote the unobserved true binary phenotype status and $\G=(G_1, \ldots, G_q)\trans$ be its associated baseline characteristics and genetic markers from the EHR linked biobank, which could be either multi-dimensional single nucleotide polymorphisms (SNPs) or a genetic risk score derived by weighting a number of SNPs. We simultaneously consider two types of error-prone outcomes or surrogates for $Y$ in our setup. First, suppose there are $p$-dimensinoal EHR surrogate features $\bX=(X_1, \ldots, X_p)\trans$ such as counts of $Y$'s related billing codes and key laboratory results. Second, let $\Naive-Logistic$ be the chart review label from experts, taking values of $k/K$ for $k\in\{0, 1, ..., K\}$, to represent different levels of certainty regarding whether the patient has the condition $Y$. In practice, $K$ is often taken as $2$ with $\Naive-Logistic\in\{0,0.5,1\}$ representing not a case, a possible case, and a case. 

Importantly, we assume the error-prone outcomes $\Naive-Logistic$ and $\bX$ only relate to genetic markers $\G$ through $Y$, i.e., $(\bNaive-Logistic, \bX) \ci \G\mid Y$. An illustration of this assumption is provided in the directed acyclic graph (DAG) of Figure \ref{fig:dag}. In this DAG, the baseline biomarkers $\G$ first occur to affect the chance of developing the disease $Y$, then $Y$ causes the downstream hospital visits producing features $\bX$ and $\U$ in EHR, where $\U$ may encode unstructured information such as images and narrative clinical notes. Though $\U$ is not directly included as an outcome in our setup, it can affect the medical review result $\Naive-Logistic$ together with the observed and structured $\bX$.

\begin{figure}[htbp]
\centering{\includegraphics[scale=1.4]{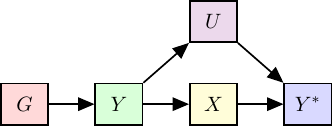}}
    \caption{\label{fig:dag} An illustrative directed acyclic graph (DAG) of the data generating mechanism.}
\end{figure}

Suppose there are $N$ patients with independent and identical copies of the complete set of variables $\bD = (Y, \Naive-Logistic,\bX, \G)$ described above, denoted as $\Dscr = \{\bD_i: i = 1,2, ...,N\}$. Since the label $\Naive-Logistic$ is derived based on expertise and additional information like $\U$, it is usually more accurate than $\bX$ in characterizing the true $Y$. However, it may still have a moderate measurement error due to incomplete information collection for medical review or complication and ambiguity of certain phenotypes. Thus, we assume that $\Naive-Logistic$ is only observable in a small set of $n$ subjects indexed by $\delta=1$, and, to account for the error of $\Naive-Logistic$, it is marginally related to $Y$ through  
\begin{equation}
\Pr(\Naive-Logistic=k/K\mid Y=y)=\lambda_{yk},~\mbox{for}~k = 0,\ldots,K,~y=0,1;\quad\blambda_{y}=\{\lambda_{y1},...,\lambda_{yK}\}.
\label{model:3}
\end{equation}
Also, note that $\bX$ and $\G$ are observed for all patients and the true outcome $Y$ is not observed for any patient. So the observed data is formed as $\Oscr = \{\bO_i = (\Naive-Logistic_i\delta_i, \delta_i, \bX_i, \G_i): i = 1,2, ...,N\}$, with the labeling indicator $\delta\perp\bD$, i.e., being completely at random. Without loss of generality, we let $\delta_i = I(1 \le i \le n)$ where $n<N$ is the size of sample with labels and $I(\cdot)$ is the indicator function. Our primary goal is to derive a risk model of $Y$ against $\G$ as well as inference of its encoded genetic associations. Since the genetic effects are usually moderate or small, it is more favorable to model and interpret $Y\sim \bG$ with a simple and parametric form to ensure good interpretability and control the estimation uncertainty. In specific, we consider a {\em working} logistic model:
\begin{align}
& \Pr(Y=1\mid \G)=g(\bbeta\trans\G), \label{model:2}
\end{align} 
where the expit link $g(x) = e^{x}/(1+e^x)$. Note that model (\ref{model:2}) is allowed to be misspecified, and we define the target model parameter as $\bbetabar = \argmax{\bbeta}\EE\ell(Y, \bbeta\trans\G)$ where $\ell(y,w)= y \log\{g(w)\} + (1-y)\log\{[1-g(w)]\}$ is the log-likelihood function of logistic regression. Though (\ref{model:2}) may be misspecified, such $\bbetabar$ is still identifiable and effective in characterizing the genetic associations. Our secondary goal is EHR phenotyping for the unobserved $Y$ using $\bX$ be deriving a risk score $\alpha(\bX)$, as well as validating its classification performance. Due to the absence of $Y$ in our observation, all above-introduced tasks are unsupervised and, thus, more challenging than the standard supervised or recent semi-supervised scenarios reviewed in Section \ref{sec:intro:review}.

\begin{remark}
Though both $\Naive-Logistic$ and $\bX$ are surrogates of the truth $Y$ with errors, we still notate and consider them separately for several reasons. First, $\Naive-Logistic$ is not accessible for a (large) fraction of subjects so the phenotyping score of $Y$ can only include the fully observed $\bX$ as the predictors and formulated as $\alpha(\bX)$. Second, although $\Naive-Logistic$ is neither perfect nor scalable, it is supposed to be more accurate and informative than $\bX$. Thus, as will be discussed in Sections \ref{sec:method} and \ref{sec:thm}, $\Naive-Logistic$ is important under our framework to stable training and efficient estimation, especially when $\bX$ is of poor quality in characterizing $Y$.  
\end{remark}

\subsection{Related literature and our contribution}\label{sec:intro:review}

Surrogate outcomes play an important role in data-driven biomedical research, particularly when obtaining the primary or true outcome of interest is costly or even impossible, e.g., demanding extensive human labor or long periods of follow-up. There is rich literature in both semi-supervised and unsupervised statistical learning with surrogates. For example, \cite{athey2019surrogate} leveraged surrogates collected in observational studies to assist learning with experimental studies in paucity of the gold standard labels. \cite{kallus2020role} and \cite{hou2021efficient} studied how to utilize surrogates to improve the efficiency of causal inference without incurring bias. \cite{hou2023risk} developed a semiparametric transformation approach to incorporate time-to-event surrogates and improve the learning efficiency with the true outcomes. 

The aforementioned literature considered a semi-supervised setting with a small sample of the true outcome $Y$. Differently, our problem setup does not involve any observation of $Y$. For such an unsupervised setting, \cite{huang2018pie} and \cite{hong2019semi} proposed maximum likelihood approaches based on parametric assumptions on the conditional model of $Y$, which enables the identification and estimation of the model coefficients. \cite{zhang2019electronic} developed a method for the unsupervised learning and phenotype validation with anchor-positive surrogate outcomes in EHR. All these recent methods largely rely on parametric model assumptions like (\ref{model:2}), a working assumption in our setup. Its misspecification could lead to biased estimation for the target parameter $\bbetabar = \argmax{\bbeta}\EE\ell(Y, \bbeta\trans\G)$ due to the absence of the true label $Y$.

Meanwhile, we notice some fully nonparametric approaches for the so called latent-structure or mixture model related to our problem setup in recent literature, including \cite{bonhomme2016estimating}, \cite{yu2019maximum}, and \cite{zheng2019nonparametric}. For example, \cite{zheng2019nonparametric} proposed a novel tensor approach for learning of nonparametric mixtures, with a key idea of introducing basis approximation to the component density functions. This track of work is in general free from the model misspecification issue discussed above but cannot provide desirable $n^{-1/2}$-consistent estimators and may encounter the ``curse of dimensionality" for multivariate surrogate outcomes.

To address the above-introduced dilemma between the bias caused by model misspecification and the low efficiency due to curse of dimensionality, we develop a {\bf T}hree-stage {\bf U}nsupervised learning approach for {\bf B}iomarkers linked with {\bf E}rror-prone outcomes, abbreviated as TUBE. Our approach primarily aims at risk modeling with the baseline biomarkers, and is also able to produce and validate a predictive EHR phenotyping score without observations of the true disease outcome. It is a semiparametric method that starts from a composite and nonparametric regression step for $\bX,Y^*$ against $\bG$ that is free of any parametric assumptions. Following this step, TUBE combines multiple surrogates for EHR phenotyping and validation, and then implements a parametric projection step to improve the interpretability and estimation efficiency of the genetic risk model. We will show that our estimator for $\bbeta$ is $n^{-1/2}$-consistent and asymptotic normal without requiring model (\ref{model:2}) to be correctly specified or $Y\sim\bX$ to have a parametric form, which are imposed by existing methods like \cite{hong2019semi} and \cite{zhang2019electronic}. Also, TUBE demonstrates significantly better performance than existing methods in our simulation and real-world studies.

\def\logit{\mbox{logit}}
\def\Sscbar{\bar{\Ssc}}
\def\sumin{\sum_{i=1}^n}
\def\sumiN{\sum_{i=1}^N}
\def\mubar{\bar{\mu}}

\section{Three-stage unsupervised learning method}\label{sec:method}

\subsection{Overview of the modeling strategy}\label{sec:method:over}

Our proposed TUBE method consists of three main steps. In stage I, we adopt an under-smoothed nonparametric and composite likelihood strategy that is free of any parametric or model structural assumptions on the forms of $Y\sim Y^*$, $Y\sim\bX$ and $Y\sim\G$. This is to avoid the potential bias caused by model misspecification on linking the error-prone outcomes $(Y^*,\bX)$ with $\G$ without the supervision of the true label $Y$. In stage II, we leverage the results from I to condense the EHR features $\bX$ into a risk score $\widehat\alpha(\bX)$ for more accurate phenotyping of $Y$, and refit the data using nonparametric likelihoods to evaluate its ROC. In stage III, we rely on the imputation outcomes from II to derive a parametric logistic model for $Y\sim\G$. Compared to the previous steps, III will output a more efficient characterization of the genetic risk or association with good interpretability and desirable convergence rates. Meanwhile, built upon previous steps robust to model misspecification, stage III will be valid even when the target genetic model is wrong.

Denote by $\mu = \Pr(Y = 1)$ and $m_{j}(x) = \Pr(Y=1 \mid X_j = x)$ for $j=1,2,\ldots,p$. To get rid of the curse of dimensionality in modeling $Y$ jointly against $X_1,X_2,\ldots,X_p$ through a multivariate nonparametric model, we consider a {\em working} conditional independence assumption across $X_1,X_2,\ldots,X_p$ given $Y$, implying an additive logistic form of their joint model:
\begin{equation}
\Pr(Y=1\mid \bX)= g\{a+\alphabar(\bX)\} \quad \mbox{with}\quad \alphabar(\bX) = \sumjp g^{-1}\{m_{j}(X_j)\},
\label{model:1}
\end{equation}
where $a$ is an intercept term introduced such that $\EE g\{\alphabar(\bX)\}=\mu$. As will be introduced in Section \ref{sec:method:stage:1}, under this construction, we can model each $X_j$ with $\bG$ separately and combine them with a composite likelihood to estimate $m_{j}(\cdot)$'s, {\em as if} $X_1\perp X_2\perp\ldots \perp X_p\mid Y$. Then we will ensemble the estimators of $m_{j}(X_j)$ through (\ref{model:1}) to derive an estimate for the phenotyping score $\alphabar(\bX)$. As we will discuss later, due to our use of the composite likelihood, violation of the additive model (\ref{model:1}) will not cause invalidity to the downstream results.

For the genetic variants $\G$, we will consider two scenarios, including that (i) $\G$ contains multi-dimensional discrete SNPs features ranging over $\{0,1,2\}$; and (2) $\G$ is a univariate continuous gene risk score. For (i), we introduce the categorical functions covering all the possible combinations of the discrete SNPs in $\G$ while for (ii), we use the spline (sieve) basis functions of $\G$. In both cases, we specify the nonparametric model of $Y\sim \G$ as 
\begin{equation}
\Pr(Y=1\mid \G)=g\{\bxi\trans\bpsi(\bG)\},
\label{equ:nonp:g}
\end{equation}
where $\bpsi(\bG)=\{\psi_1(\bG),\psi_2(\bG),\ldots,\psi_{d_g}(\bG)\}\trans$ is a set of bases with possibly diverging dimensionality, used to approximate any (smooth) functions of $\bG$. Note that model (\ref{equ:nonp:g}) is a {\em nuisance} model introduced to avoid model misspecification in the first stage of our method. Our final goal is to estimate the parametric model (\ref{model:2}) with a more desirable convergence rate as well as easier interpretation than (\ref{equ:nonp:g}). This is more advantagous especially when the genetic association is mild or small and, thus, requiring small enough estimation uncertainty to detect.

\subsection{Stage I: sieve-approximated composite likelihood}\label{sec:method:stage:1}

We first focus on the estimation of $m_j(\cdot)$'s and $\Pr(Y=1\mid \G)$. To ensure the validity while incorporating the additional genetic information, we consider a composite log-likelihood formulated under our key assumption that $(\bNaive-Logistic, \bX) \ci \G\mid Y$ and a {\em working} independence condition of $X_1,...,X_p$ given $Y$:
\[
\sumin\log \left\{\sum_{y=0}^1 \Pr(Y^*_{i}\mid Y_i=y) \Pr(Y_i=y\mid \bG_i) \right\}+\sumiN\sumjp \log \left\{\sum_{y=0}^1 \frac{\Pr(Y_i=y\mid X_{ij}) \Pr(Y_i=y\mid \bG_i)}{\Pr(Y_i=y)}\right\},
\]
where $X_{ij}$ is the $j$-th EHR outcome of subject $i$. As is outlined in Section \ref{sec:method:over}, due to potential misspecification of the parametric models like (\ref{model:2}), we model $\Pr(Y=y\mid \bG)$ nonparametrically by (\ref{equ:nonp:g}), and adopt a similar sieve construction on each
\[
m_j(X_j)=\Pr(Y=1\mid X_{j})=g\{\bzeta_j\trans\bvarphi_j(X_{j})\},
\]
where $\bvarphi_j(x)$ is a vector of basis functions used to approximate $g^{-1}\{m_{j}(x)\}$. For discrete $X_j$, we naturally set $\bvarphi_j(x)$ as its dummy variables. For continuous $X_j$, we again use sieve.
Then we can construct the sieve-approximated composite likelihood as:
\begin{equation*}
\Csc(\btheta)=\sum_{i=1}^n\log\left(\sum_{y=0}^1 \lambda_{yY_{i}^*}g_y\{\bxi\trans\bpsi(\bG_i)\} \right)+\sumiN\sumjp \log \left(\sum_{y=0}^1 \mu_y^{-1}g_y\{\bzeta_j\trans\bvarphi_j(X_{ij})\}g_y\{\bxi\trans\bpsi(\bG_i)\}\right),
\end{equation*}
where $\btheta=\{\bxi,\bzeta,\blambda,\mu\}$, $\blambda=(\blambda_0\trans,\blambda_1\trans)\trans$, $\bzeta=(\bzeta_1\trans,\ldots,\bzeta_p\trans)\trans$, and we denote by $g_y(\cdot)=y g(\cdot)+(1-y)\{1-g(\cdot)\}$ and $\mu_y=\Pr(Y=y)=y\mu+(1-y)(1-\mu)$. To solve for $\btheta$ that maximizes $\Csc(\btheta)$, we propose to use an expectation???maximization (EM) algorithm outlined in Algorithm \ref{alg:1}.

\begin{algorithm}[H]
\caption{\label{alg:1} EM algorithm for the nonparametric composite log-likelihood.}
{\bf Input:} Observed data $\Oscr = \{\bO_i = (\Naive-Logistic_i\delta_i, \delta_i, \bX_i, \G_i): i = 1,2, ...,N\}$.
 ~\\
{\bf Initialize} with $\bthetahat^{(0)}=\{\widehat\bxi^{(0)},\widehat\bzeta^{(0)},\widehat\blambda^{(0)},\widehat\mu^{(0)}\}$ obtained by Algorithm \ref{alg:app:1}. Iterate on the following two steps for $r=0,1,\ldots,R$ until convergence.
 ~\\
{\bf E-step}. For each subject $i$ and outcome $j$ (or $Y^*$ if observed: $\delta_i=1$), impute the probability for the unobserved $Y_i$ conditional on the covariates in each component of the composite likelihood:
\[
\widehat Y_{i0}^{(r+1)}=\delta_i\times\frac{\widehat\lambda_{1Y_{i}^*}^{(r)}g_1\{\bpsi\trans(\bG_i)\widehat\bxi^{(r)}\}}{\sum_{y=0}^1 \widehat\lambda_{yY_{i}^*}^{(r)}g_y\{\bpsi\trans(\bG_i)\widehat\bxi^{(r)}\}};~\widehat Y_{ij}^{(r+1)}=\frac{g_1\{\bvarphi\trans_j(X_{ij})\widehat\bzeta^{(r)}_j\}g_1\{\bpsi\trans(\bG_i)\widehat\bxi^{(r)}\}/\widehat\mu_1^{(r)}}{\sum_{y=0}^1g_y\{\bvarphi\trans_j(X_{ij})\widehat\bzeta^{(r)}_j\}g_y\{\bpsi\trans(\bG_i)\widehat\bxi^{(r)}\}/\widehat\mu_y^{(r)}}.
\]
 ~\\
{\bf M-step}. Update $\btheta$ through the maximum likelihood estimation (MLE) specified with the imputed outcomes from the E-step: 
\begin{align*}
&\widehat\mu^{(r+1)}=\frac{1}{Np+n}\sumiN\sum_{j=0}^p \widehat Y_{ij}^{(r+1)};\quad\widehat\lambda_{yk}^{(r+1)}= \frac{\sum_{i=1}^n  I (Y^*_{i}= k) \{\Yhat_{i0}^{(r+1)}\}^y \{1-\Yhat_{i0}^{(r+1)}\}^{1-y}} {\sum_{i=1}^n\{\Yhat_{i0}^{(r+1)}\}^y \{1-\Yhat_{i0}^{(r+1)}\}^{1-y}};\\
&\widehat\bxi^{(r+1)}=\argmax{\bxi}\sum_{i=1}^n \ell\left(\Yhat_{i0}^{(r+1)},\bpsi\trans(\bG_i)\bxi\right)+ \sumiN\sumjp \ell\left(\widehat Y_{ij}^{(r+1)}, \bpsi\trans(\bG_i)\bxi\right);\\
&\widehat\bzeta_j^{(r+1)}=\argmax{\bzeta_j}\sumiN \ell\left(\widehat Y_{ij}^{(r+1)}, \bvarphi\trans_j(X_{ij})\bzeta_j\right),\quad\mbox{for }j=1,2,\ldots,p.
\end{align*}
 ~\\
{\bf Output:} $\bthetahat=\{\widehat\bxi,\widehat\bzeta,\widehat\blambda,\widehat\mu\}=\{\widehat\bxi^{(R)},\widehat\bzeta^{(R)},\widehat\blambda^{(R)},\widehat\mu^{(R)}\}$

\end{algorithm}

Algorithm \ref{alg:1} iterates on two main steps. First, there is an E-step imputing the unobserved true outcome $Y$ separately conditional on each $(X_j,\bG)$ or $(Y^*,\bG)$ as the set of features appearing in each component of the composite likelihood. Unlike the EM algorithms for joint likelihood objectives, our method does not involve any imputation model of $Y$ using the whole set of observed variables $(\bX,\bG,Y^*)$. This in turn ensures the validity free of any assumptions on the joint distribution of $\bX,Y^*$ that is hard to characterize due to the curse of dimensionality. Second, Algorithm \ref{alg:1} involves an M-step solving for $\btheta$ through MLE constructed using the imputed $Y$'s. Again, corresponding to the composite likelihood construction, $\blambda$ and $\bzeta_j$'s for different error-prone outcomes are solved separately based on their own imputed outcomes. 

In Theorem \ref{thm:1} presented later, we show that Algorithm \ref{alg:1} maintains an ascent property on the objective composite likelihood function that is desirable for optimization. Nevertheless, it is still practically crucial to have a good initial estimator $\bthetahat^{(0)}$ for Algorithm \ref{alg:1} to avoid the local minima issue. In response to this, we propose in Algorithm \ref{alg:app:1} of Appendix to derive $\widehat\bxi^{(0)},\widehat\bzeta^{(0)},\widehat\mu^{(0)}$ through MLE constructed as if $I(Y^*=1)$ was the true outcome, i.e., the logistic regression of $I(Y^*=1)$ against $\bpsi(\bG)$ or each $\bvarphi_j(X_{j})$. For $\widehat\blambda^{(0)}$, we set it up with a proper guess presuming that $Y^*$ is informative.

\subsection{Stage II: condensing EHR features for phenotyping}

With the fitted estimator in Stage I, we derive $\alphahat(\bX)=\sumjp \bvarphi\trans_j(X_{j})\widehat\bzeta_j$, serving as a phenotype score condensing the outcomes $X_1,X_2,\ldots,X_p$. For $\alphahat(\bX)$, we further adopt a nonparametric likelihood approach that combines it with $\bG$ to derive an imputation model for $Y$. Since $\alphahat(\bX)$ ensembles multiple EHR outcomes, it tends to be more predictive of $Y$ than each single $m_j(X_j)$. So this procedure can be more efficient than modeling each single $X_j$ separately in $\Csc(\btheta)$, thus, being more favorable for the downstream analysis. As implied by (\ref{model:1}), the optimal ensemble is $\alphabar(\bX) = \sumjp g^{-1}\{m_{j}(X_j)\}$ only when the working assumption $X_1\perp X_2\perp\ldots \perp X_p\mid Y$ holds. When there is a strong evidence that such conditional independence does not hold, an alternative strategy is to set the phenotyping score $\alpha(\bX)$ as the first principle component of $g^{-1}\{m_{j}(X_j)\}$ for $j=1,2,\ldots,p$, to make it representative of the multiple EHR outcomes.

Again, we will not rely on any parametric or model structural assumptions on the sensitivity function $\Tsc\subalpbary(c)=\Pr(\alphabar(\bX)> c\mid Y=y)$ for $c\in\mathbb{R}$ and $y\in\{0,1\}$ that captures $\alphabar(\bX)\mid Y$. In this case, the log-likelihood function can be written as 
\[
\sum_{i=1}^n\log \left\{\sum_{y=0}^1\lambda_{yY^*_{i}} g_y\{\bxi\trans\bpsi(\bG_i)\}\right\}+\sumiN \log\left\{ -\sum_{y=0}^1\dot{\Tsc}\subalphaty\{\alphahat(\bX_i)\} g_y\{\bxi\trans\bpsi(\bG_i)\}\right\}.
\]
Without any further constraint on $\Tsc\subalpbary(c)=\Pr(\alphabar(\bX)> c\mid Y=y)$, the above log-likelihood function will not have a unique maximizer. Thus, inspired by existing literature in nonparametric MLE \citep[e.g.]{murphy2000profile}, we restrict $\Tsc\subalpbary(c)$ to be a step function that can only jump at the observed data points $\{\alphahat(\bX_i):i=1,2,\ldots,N\}$, and denote its jump size at each $\alphahat(\bX_i)$ as $\nabla\Tsc\subalpbary\{\alphahat(\bX_i)\}$. If the true status $Y_i$ was observed, the MLE for $\Tsc\subalphaty(c)$ under this step-function constraint would be derived as
\[
\breve\Tsc\subalphaty(c)=\frac{\sum_{i=1}^N I(\alphahat(\bX_i)> c)I(Y_i=y)}{\sum_{i=1}^N I(Y_i=y)}\quad\mbox{for}\quad c=\alphahat(\bX_{i'}).
\]
Based on this, our objective becomes to maximize
\begin{equation}
\Lsc(\boldsymbol{\eta}_{\alphahat})
=\sum_{i=1}^n\log \left\{\sum_{y=0}^1\lambda_{yY^*_{i}} g_y\{\bxi\trans\bpsi(\bG_i)\}\right\}+\sumiN \log\left\{ \sum_{y=0}^1-\nabla\Tsc\subalphaty\{\alphahat(\bX_i)\} g_y\{\bxi\trans\bpsi(\bG_i)\}\right\},
\label{equ:prof:llh}
\end{equation}
where $\boldsymbol{\eta}_{\balpha}=\{\Tsc\subalpzero(\cdot),\Tsc\subalpone(\cdot),\blambda,\bxi\}$, under the step-function constraints on $\Tsc\subalpzero(\cdot),\Tsc\subalpone(\cdot)$. Since we do not specify the correlation or dependence between $Y^*$ and $\alphahat(\bX)$, we still adopt a composite strategy to model them in (\ref{equ:prof:llh}). But different from the fully composite $\Csc(\btheta)$ also treating $X_j$ separately, we now condense $X_j$'s into a single $\alphahat(\bX)$.

Similar to Algorithm \ref{alg:1}, we adopt an EM algorithm to numerically maximize the objective $\Lsc(\boldsymbol{\eta}_{\alphahat})$ for the solution $\widetilde{\boldsymbol{\eta}}_{\alphahat}=\{\widetilde\Tsc\subalphatzero(\cdot),\widetilde\Tsc\subalphatone(\cdot),\widetilde\blambda,\widetilde\bxi\}$; see Algorithm \ref{alg:app:1} in Appendix \ref{sec:app:impl}. At last, we introduce Theorem \ref{thm:1} to establish the ascent properties of our proposed EM algorithms for $\Csc(\btheta)$ and $\Lsc_{\alphahat}(\bfeta)$ formulated in Steps I and II respectively.
\begin{theorem}
Let $\bthetahat^{(r)}$ and $\widetilde{\bfeta}^{(r)}$ be the estimators at the $r$-th iteration of the EM Algorithms \ref{alg:1} and \ref{alg:app:2} respectively. We have $\Csc(\bthetahat^{(r)})\leq \Csc(\bthetahat^{(r+1)})$ and $\Lsc(\widetilde{\bfeta}_{\alphahat}^{(r)})\leq\Lsc(\widetilde{\bfeta}_{\alphahat}^{(r+1)})$, i.e., each iteration in our EM algorithms is ensured to result in the ascent of the objective log-likelihood functions. 
\label{thm:1}
\end{theorem}

\subsection{Stage III: genetic risk modeling and EHR phenotype validation}\label{sec:gene:ehr}

In Steps (I) and (II) introduced above, we fit nonparametric models for $Y \mid \bG$ to make the estimators $\alphahat(\cdot)$ and $\Sschat\subalpbary(\cdot)$ more robust to model misspecification. In practice, directly using such nonparametric models for gene association analysis often results in large variance or even inefficiency due to the curse of dimensionality. Thus, in this step, we leverage the extracted $\widetilde{\boldsymbol{\eta}}_{\alphahat}$ to construct a parametric genetic risk for the true outcome $Y_i$ against $\bG_i$. In specific, with $\widetilde{\boldsymbol{\eta}}_{\alphahat}$, we characterize $\EE[Y_i\mid \alphabar(\bX_i),\bG_i]$ for all $i=1,2,\ldots,N$, and $\EE[Y_i\mid \Naive-Logistic_i,\bG_i]$ for $i=1,2,\ldots,n$ as
\[
\widetilde Y_{i0}=\frac{\widetilde\lambda_{1Y_{i}^*}g_1\{\bpsi\trans(\bG_i)\widetilde\bxi\}}{\sum_{y=0}^1 \widetilde\lambda_{yY_{i}^*}^{(r)}g_y\{\bpsi\trans(\bG_i)\widetilde\bxi\}};\quad\widetilde Y_{i1}=\frac{\nabla\widetilde\Tsc\subalphatone\{\alphahat(\bX_i)\}g_1\{\bpsi\trans(\bG_i)\widetilde\bxi\}}{\sum_{y=0}^1\nabla\widetilde\Tsc\subalphaty\{\alphahat(\bX_i)\}g_y\{\bpsi\trans(\bG_i)\widetilde\bxi\}},
\]
which coincides with the imputation of the unobserved $Y$ in the last E-step of Algorithm \ref{alg:app:1}. Note that 
$\widetilde Y_{i1}$ is not necessarily consistent for $\EE[Y_i\mid \bX_i,\bG_i]$ unless the working independence assumption (\ref{model:1}) holds and $\EE[Y_i\mid \bX_i]=\EE[Y_i\mid \alphabar(\bX_i)]$. Then we conduct logistic regression for the imputed outcomes $\widetilde Y_{i0}$ and $\widetilde Y_{i1}$ separately against $\bG_i$, to obtain estimators
\begin{align*}
    \bbetatilde_0 &= \argmax{\bbeta}\sum_{i=1}^n\ell(\Ytilde_{i0}, \G_i\trans\bbeta);\quad
    \bbetatilde_1 = \argmax{\bbeta}\sumiN \ell(\Ytilde_{i1}, \G_i\trans\bbeta).
\end{align*}
Although $N > n$, the standard error of $\widetilde \bbeta_0$ may still be smaller than that of $\widetilde \bbeta_1$ since $X$ is typically less informative than the chart review labels $Y^*$ in terms of measuring the true $Y$. To derive a more efficient estimator, the final step is to assemble $\bbetatilde_0$ and $\bbetatilde_1$ as:
\[
\bbetatilde = \widehat\omega \bbetatilde_0+(1-\widehat\omega)\bbetatilde_1;\quad \widehat\omega\in[0,1],
\]
where $\widehat\omega$ is a weight determined using the data to minimize the variance of $\bbetatilde$ among all convex combinations of $\bbetatilde_0$ and $\bbetatilde_1$. When $N\gg n$, we can show that $\bbetatilde_0$ and $\bbetatilde_1$ are asymptotically independent, and, thus, the optimal weight $\widehat\omega={\widehat{\rm SE}_0^{-2}}/{(\widehat {\rm SE}_0^{-2}+\widehat{\rm SE}_1^{-2})}$, where $\widehat{\rm SE}_0$ and $\widehat{\rm SE}_1$ represent the estimated standard error of $\bbetatilde_0$ and $\bbetatilde_1$. In general, we can take 
\[
\widehat\omega=\arg\min_{\omega\in[0,1]} (\omega,1-\omega)\widehat\Sigma_{\widetilde\bbeta_0,\widetilde\bbeta_1}(\omega,1-\omega)\trans,
\]
where $\widehat\Sigma_{\widetilde\bbeta_0,\widetilde\bbeta_1}$ is the asymptotic covariance matrix of $(\bbetatilde_0,\bbetatilde_1)$ computed using bootstrap. Since the true disease status $Y$ is unobserved, the estimators $\widetilde\bbeta_0$ and $\widetilde\bbeta_1$ are subject to the issue that the switch between $Y=0$ and $Y=1$ cannot be identified from the observed data. To address this, we assume the coefficient for $G_1$ to be greater than zero with $G_1$ chosen as an informative feature to $Y$. Correspondingly, we shall flip the sign of the fitted $\bbetatilde_0$ or $\bbetatilde_1$ if $\betatilde_{01}<0$ or $\betatilde_{11}<0$. Alternatively, one could also restrict the prevalence of $Y$ to be smaller than $0.5$, which does not require the knowledge of some informative feature $G_1$. 

As the by-product, we are also able to validate the derived phenotyping score $\alphahat(\bX)$ using the fitted sensitivity functional $\widetilde\Tsc\subalphaty(\cdot)$. Denote the limiting (population-level) function of $\alphahat(\bX)$ as $\alphabar(\bX)$. The true positive rate (TPR) and false positive rate (FPR) of the classifier $I(\widehat\alpha(\bX)>c)$ or $I(\alphabar(\bX)>c)$ on the true label $Y$ can be naturally estimated using  $\widetilde\Tsc\subalphatone(c)$ and $\widetilde\Tsc\subalphatzero(c)$ respectively. Furthermore, the receiver operating characteristic (ROC) curve of $\alphahat(\bX)$ or $\alphabar(\bX)$ can be estimated by $\widehat\roc(u) = \widetilde\Ssc\subalphatone\{\widetilde\Ssc^{-1}\subalphatzero(u)\}$ for $u\in[0,1]$, and the area under ROC $\widehat\auc = \int_0^1 \widehat\roc(u) du$.

\section{Asymptotic analysis}\label{sec:thm}

In this section, we provide asymptotic analysis of the TUBE estimators $\alphahat(\bX)$, $\widetilde\Ssc\subpiy(\cdot)$, and  $\widetilde\bbeta$ resulted from our described steps in Sections \ref{sec:method:stage:1}--\ref{sec:gene:ehr}. We consider $\G$ as a continuous univariate gene risk score and $\psi(\G)$ as its spline basis function. Let $\bthetabar=\{\bar{\bxi},\bar{\bzeta},\bar{\blambda},\mubar\}$ and $\bar{\boldsymbol{\eta}} = \{\bar{\Ssc}_{\alphabar,1},\bar{\Ssc}_{\alphabar,0}, \bar{\blambda}, \bar{\bxi}\}$ be the population-level (true) parameters. We define the norm of $\btheta$ to be $\|\btheta\|_2 = \left\{\EE\{\|\bxi\|_2^2\} + \EE\{\|\bzeta\|_2^2\}+ \EE\{\|\blambda\|_2^2\} + \EE\{u^2\}\right\}^{1/2}$ and the norm of $\boldsymbol{\eta}$ to be $\|\boldsymbol{\eta}\|_2 = \left\{\sum_{y=0}^1\int (\Ssc_{\alpha, y}(c))^2dc + \EE\{\|\blambda\|_2^2\} + \EE\{\|\bxi\|_2^2\}\right\}^{1/2}$. We first introduce smoothness and regularity assumptions as follows.

\begin{assume}\label{asu:1}
Covariates $(\bX,\G)$ have compact domain $\mathcal{X}\times \mathcal{G}$ with their joint probability density function being twice continuously differentiable. For all $j=1,2,\ldots,p$ and $y=0,1$, $m_{jy}(x)$ and $\gamma_{y}(g)$ are twice continuously differentiable. For $y=0,1$, $\Ssc'\subpiy(c)$, the derivative of $\Ssc\subpiy(c)$ is continuously differentiable. 
\end{assume}

\begin{assume}\label{asu:4}
The parameter spaces of $\bar{\boldsymbol{\theta}}$ and $\bar{\boldsymbol{\eta}}$ are compact. Hessian matrix $\EE[\G\G\trans g_1'(\G\trans\bbeta_0)]$ has its all eigenvalues staying away from $0$ and $\infty$. For any $\btheta_1,\btheta_2$ and $\boldsymbol{\eta_1},\boldsymbol{\eta_2}$, $\EE[\Csc(\btheta_1+\tau(\btheta_2-\btheta_1))]$ and $\EE[\Lsc(\bfeta_{\alpha,1}+\tau(\bfeta_{\alpha,2}-\bfeta_{\alpha,1}))]$ are twice continuously differentiable with respect to $\tau\in[0,1]$, $\frac{\partial^2}{\partial\tau^2}\EE[\Csc(\btheta_1+\tau(\btheta_2-\btheta_1))]\asymp-\|\btheta_2-\btheta_1\|_2^2$, and $\frac{\partial^2}{\partial\tau^2}\EE[\Lsc(\bfeta_{\alpha,1}+\tau(\bfeta_{2}-\bfeta_{1}))]\asymp-\|\bfeta_{\alpha,2}-\bfeta_{\alpha,1}\|_2^2$. 
\end{assume}

\begin{remark}
Assumption \ref{asu:1} consists of mild smoothness conditions commonly used for the asymptotic analysis of of M-estimation and sieve-smoothed regression \citep[e.g.]{van2000asymptotic,CHEN20075549}. Assumption \ref{asu:4} requires the non-singularity of the hessian matrix as well as the strong convexity of the loss functions, which has been also frequently used in the literature. 
\end{remark}

\begin{remark}
When $\bX$ and $\G$ are discrete, e.g., $\G$ being the categorical functions of several SNPs, Assumption \ref{asu:1} will be as given. In such a situation with discrete $\bX$, the sensitivity function $\Ssc\subpiy(c)$ will only have finite choices on the cutoff $c$, and the asymptotic analysis of its estimator will be degenerated and simplified.     
\end{remark}

Next, we establish the consistency and asymptotic normality for the phenotyping score $\alphahat(\bx)$ in Theorem \ref{thm:2}, as well as those for the estimator of its sensitivity function in Theorem \ref{thm:3}. Let $J_{N}$ be the dimensionality of the bases $\bvarphi_j(\bX)$ and $\bpsi(\G)$ supposed to increase with $N$.

\begin{theorem}
Under Assumptions \ref{asu:1} and \ref{asu:4} and assume that $N^{1/4}\ll J_{N}\ll N^{1/2}$. As $n,N\rightarrow\infty$, $\sup_{\bx\in \mathcal{X}}|\alphahat(\bx) - \alphabar(\bx)|$ converges to $0$ in probability. Moreover, for $\bx\in\mathcal{X}$, $\sqrt{{N}/{J_N}}\{\alphahat(\bx) - \alphabar(\bx)\}$ converges weakly to some zero-mean Gaussian process. 
\label{thm:2}
\end{theorem}

\begin{theorem}
Under all assumptions in Theorem \ref{thm:2}, then as $n,N \to \infty$, $\sup_{c\in\mathbb{R}}|\widetilde\Tsc_{\alphahat,0}(c) - \bar{\Tsc}_{\alphabar,0}(c)|+|\widetilde\Tsc_{\alphahat,1}(c) - \bar{\Tsc}_{\alphabar,1}(c)|$ converges to $0$ in probability, and for $c\in \mathbb{R}$, $\sqrt{{N}/{J_N}}\{\widetilde\Tsc_{\alphahat,0}(c) - \bar{\Tsc}_{\alphabar,0}(c),\widetilde\Tsc_{\alphahat,0}(c) - \bar{\Tsc}_{\alphabar,0}(c)\}$ converges weakly to some zero-mean Gaussian process for $c\in\mathbb{R}$. 

\label{thm:3}
\end{theorem}

Considering that our primary goal is the genetic risk estimation with $\widetilde \bbeta$, we under-smooth the sieve estimator of $\alphabar$ by taking $J_{N}$ slightly larger than $O(N^{1/4})$, to achieve the asymptotic unbiasedness and normality of $\widetilde \bbeta$ that will be established in Theorem \ref{thm:4}. This choice of $J_{N}$ does not lead to the optimal convergence rate of these by-products $\alphahat(\bx)$ and $\widetilde\Tsc_{\alphahat,y}(c)$. To further refine these estimators, one just needs to take $J_{N} \asymp N^{1/5}$ and carry out Steps I and II. This leads to the $N^{-2/5}$-convergence of $\alphahat(\bx) - \alphabar(\bx)$ and $\widetilde\Tsc_{\alphahat,y}(c) - \bar{\Tsc}_{\alphabar,y}(c)$, an improvement compared to the current $N^{-3/8}$-convergence. However, the estimator derived with $J_{N} \asymp N^{1/5}$ cannot ensure the desirable parametric rate and asymptotic normality of $\widetilde \bbeta_0$ and $\widetilde \bbeta_1$ obtained in Step III. See existing literature like \cite{CHEN20075549} for more relevant results.

Finally, we establish the convergence properties of $\widetilde \bbeta_0$ and $\widetilde \bbeta_1$, which reveals the $n^{1/2}$-consistency and asymptotic normality of the TUBE estimator $\widetilde\bbeta$.

\begin{theorem}
Under all assumptions in Theorem \ref{thm:2}, both $\widetilde \bbeta_0$ and $\widetilde\bbeta_1$ converge to $\bar{\bbeta}$ in probability and $\{\sqrt{n}(\widetilde \bbeta_0 - \bar{\bbeta}),\sqrt{N}(\widetilde \bbeta_1 - \bar{\bbeta})\}$ converges weakly to a zero-mean Gaussian distribution.
\label{thm:4}
\end{theorem}

\section{Simulation}\label{sec:simu}

We conduct comprehensive simulation studies to evaluate the finite-sample performance of the proposed method. Let \textrm{Binomial}$\left\{n, p\right\}$ denote the binomial distribution with $n$ trials and a success probability of $p$. To generate risk factors $\G=(G_1, \ldots, G_q)\trans$, we consider $q=4$ with $G_1\sim{\rm N}(0, 1)$, and $G_2$, $G_3$, $G_4$ generated independently from \textrm{Binomial}$\left\{2, 0.6\right\}$. For generation of the unobserved true outcome $Y$ and EHR surrogates $\bX$, we consider the following three settings:
\begin{itemize}
    \item[(a)] $Y\sim\textrm{Bernoulli}\left\{g(\bf G\trans\bbeta)\right\}$ where $\bbeta^*=(-4.6,1.6,1.6,1.6,1.6)\trans$; and $\bX=\{Y+0.5(1-Y)+\epsilon_1, Y+0.5(1-Y)+\epsilon_2, 0.5Y+0.25(1-Y)+\epsilon_3\}\trans$ where $\epsilon_1,\epsilon_2$, $\epsilon_3$ are independent standard normal noises.
    
    \item[(b)] $Y\sim\textrm{Bernoulli}\left\{g(G_1+G_1^2-\cos(G_1)-G_2-G_3-G_4+2)\right\}$, with $\bX$ generated given $Y$ in the same way as (a).
    
    \item[(c)] $Y\sim\textrm{Bernoulli}\left\{g(-G_1+G_1^2+\sin (G_1)-G_2-G_3-G_4+1)\right\}$; and $\bX=\{Y+0.5(1-Y)+0.005G_1+\epsilon_1, Y+0.5(1-Y)+0.005G_1+\epsilon_2, 0.5Y+0.25(1-Y)+0.005G_1+\epsilon_3\}\trans$ where $\epsilon_1,\epsilon_2$, $\epsilon_3$ are independent standard normal noises.
\end{itemize}
In all settings, we set $N = 10000$ and generate $Y^*$ from $\textrm{Binomial}\left\{2, \textrm{expit}(-2+4Y+0.1_3\trans \bX)\right\}$. As discussed earlier, $Y^*$ is supposed to be an imperfect but more informative outcome compared to $\bX$. Our setup mimics this by imposing a much stronger effect of $Y$ on $Y^*$. We also let the size of $Y^*$ labels $n$ range from $100$ to $1000$ to investigate its influence on the efficiency of the methods.

We consider the following three methods for comparison: (1) the simple approach referred as Naive-Logistic directly using the label $Y^*$ as the outcome for analysis; (2) our main benchmark \cite{hong2019semi} using the composite likelihood approach with parametric modeling on $\bX$ and $\bG$; (3) the proposed TUBE approach with $\bpsi(\bG) = (\bpsi_1(G_1), G_2, G_3, G_4)$ and the basis functions $\bvarphi_j$ and $\bpsi_1(G_1)$ specified as the natural spline with the degree of freedom as $4$. Note that \cite{hong2019semi}'s method is fully parametric and, thus, will concur the issues of model misspecification in settings (b) and (c) due to the non-linearity of $Y\sim\bG$. In setting (c), we introduce some small indirect effect of $\bG$ on $\bX$ given $Y$ that moderately breaks our key independence assumption $\bX\perp\bG\mid Y$. This is to examine the sensitivity to the (slight) violation of this assumption.

The parameters of our interests include $\bbeta$, the logistic model coefficients obtained by regressing $Y$ against $\bG$, as well as the accuracy parameter AUC of $Y$ against their phenotyping score obtained in each method. The population level parameters of $\bbeta$ and $\bG$ are computed by generating an extremely large sample. Our evaluation metrics include mean squared error (MSE) in Figure \ref{fig:sim:2}, percent bias in Figure \ref{fig:sim:1}, i.e., the ratio between absolute bias and root MSE, and coverage probability (CP) of the 95\% CI computed using the standard resampling bootstrap procedure; see Figure \ref{fig:sim:3}. The results in Figures \ref{fig:sim:2}-\ref{fig:sim:3} are obtained based on $500$ times of simulation. For the multi-dimensional $\bbeta$, we only present the average performance over $\beta_1,\ldots,\beta_4$ in these figures and the element-wise results can be found in the tables of Appendix \ref{sec:app:sim}.


\begin{figure}[H]
\centering{\includegraphics[width=0.8\textwidth]{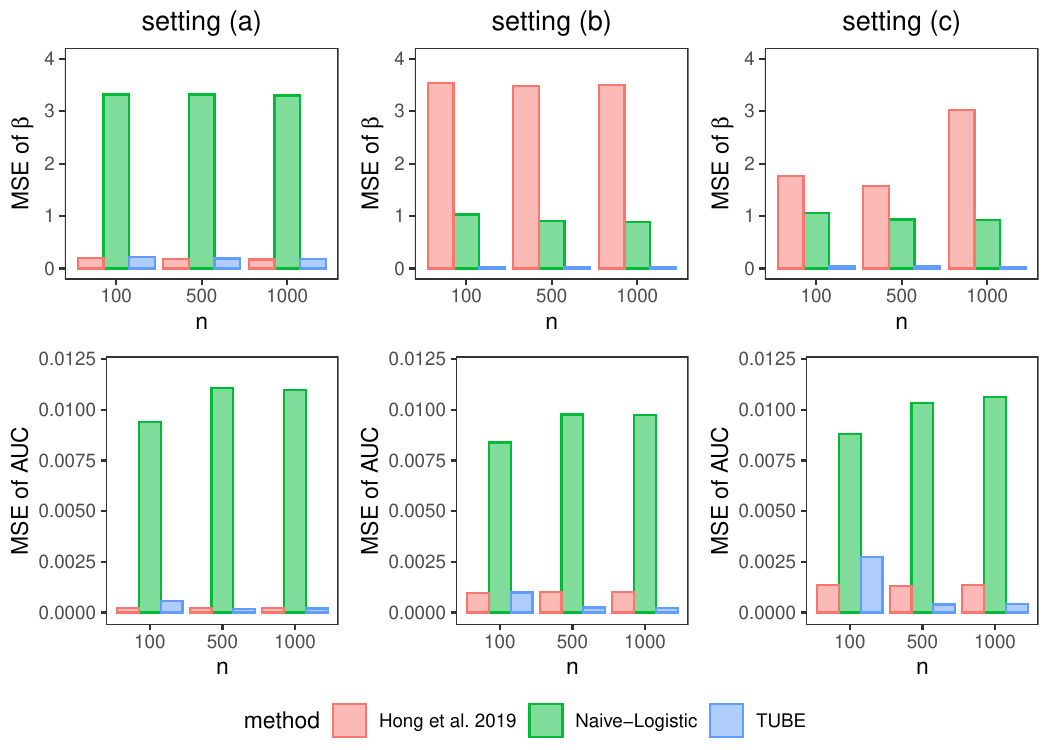}}
    \caption{Mean squared error (MSE) for estimators of the genetic effects $\bbeta$ and the AUC of the phenotyping score in different settings introduced in Section \ref{sec:simu}.}
    \label{fig:sim:2}
\end{figure}

\begin{figure}[H]
\centering{\includegraphics[width=0.8\textwidth]{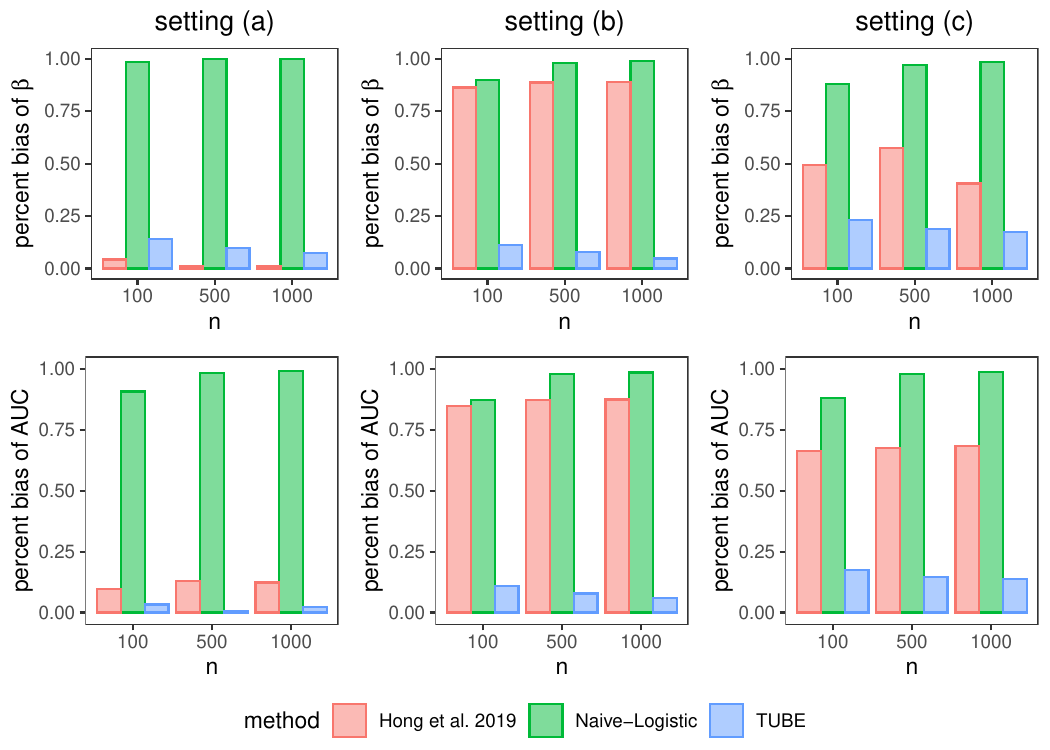}}\\
    \caption{Absolute Biases/RMSE for estimators of the genetic effects $\bbeta$ and the AUC of the phenotyping score in different settings introduced in Section \ref{sec:simu}.}
    \label{fig:sim:1}
\end{figure}

\begin{figure}[H]
\centering{\includegraphics[width=0.8\textwidth]{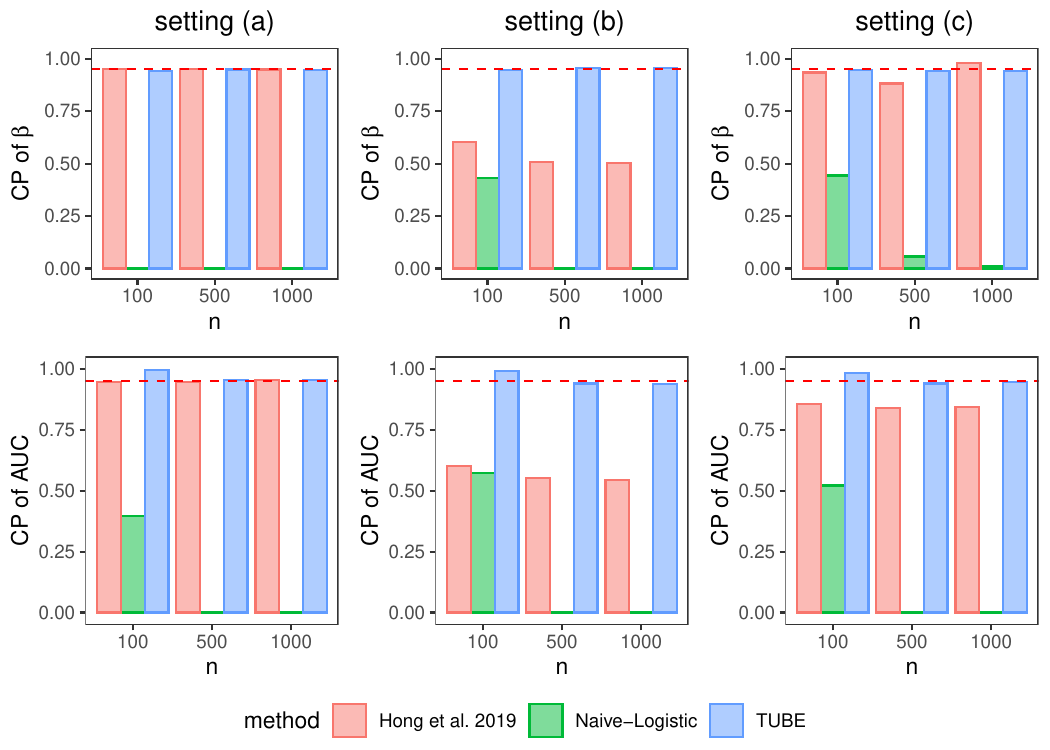}}\\
    \caption{Coverage probabilities (CP) for estimators of the genetic effects $\bbeta$ and the AUC of the phenotyping score in different settings introduced in Section \ref{sec:simu}.}
    \label{fig:sim:3}
\end{figure}

In all settings, Naive-Logistic shows large MSEs and percent biases due to the erroneousness of $Y^*$ in measuring the true $Y$. In setting (a), TUBE attains close performance to the benchmark methods in \cite{hong2019semi} that relies on a fully parametric modeling strategy and does not encounter the model misspecification issue. In specific, the percentage difference in the MSE between the two methods is smaller than $5\%$ on all parameters when $n\geq 500$ in setting (a). Also, both methods attain small enough percent bias and desirable coverage probability on $\bbeta$ and AUC. Thus, although it seems redundant to use a more complex semiparametric modeling strategy in TUBE compared to  \cite{hong2019semi} when the true models are indeed linear and parametric, this complexity does not result in TUBE's loss of validity or efficiency. This result is in line with our conclusions in Section \ref{sec:thm} that the sieve estimators does not impact the parametric rate of our estimator for $\bbeta$ due to under-smoothing.

In settings (b) and (c) under which the fully parametric method of \cite{hong2019semi} has a severe issue in model misspecification, TUBE achieves significantly better performance than \cite{hong2019semi} and ensures the validity of inference. For example, under setting (b) with $n=500$, the average MSE of TUBE on $\bbeta$ is more than 90\% smaller than that of \cite{hong2019semi}. Also, TUBE successfully maintains a small percent bias (5\%--10\%) and appropriate coverage probability while \cite{hong2019semi} fails to provide valid inference with the average coverage rates around 30\% below than the nominal level 95\% in setting (b). This substantial improvement of TUBE is resulted from the nonparametric construction in our Steps I and II that protect our approach against bias due to the nonlinear effects.

In addition, we notice that as the labeled sample size $n$ increases, the MSEs of TUBE on $\bbeta$ and AUC gradually decrease as $Y^*$ provides additional information over $\bX$. For example, when $n$ increase from $100$ to $500$, TUBE's MSE on AUC decreases more than $50\%$ in all settings. Recall that in practice and our simulation setup, $Y^*$ is usually more informative than $\bX$ even though both of them contains errors in measuring the true $Y$. Thus, moderately increasing the size of $Y^*$ could result in efficiency gain even with the total sample size $N$ unchanged. Meanwhile, we do not see the improvement of Naive-Logistic and \cite{hong2019semi} as $n$ increases in settings (b) and (c) probably because of their large bias.

\section{Real Example}\label{sec:real}

The rising incidence of Type II diabetes mellitus (T2D) in recent years has risen great concern in health. Previous genome-wide association studies (GWAS) have identified many genetic variations associated with insulin resistance or inadequate insulin production attributing to T2D \citep{mahajan2018fine}. Consequently, polygenic risk score (GRS) has been developed to predict individual's genetic risk of developing T2D \citep{he2021comparisons}. These advancements provide great potential for precision medicine approaches in the prevention and management of the T2D disease. In this application, we study the Mass General Brigham (MGB) biobank data \citep{castro2022mass} with a primary goal to build a genetic risk prediction model for T2D using its GRS and demographic information.

Our data set includes $N=16,963$ MGB biobank participants up to 2021 with their available EHR features updated for the same year. Their risk factors $\bG$ contain $G_1$, an one-dimensional {\em GRS} for T2D derived using the reported variants and effect sizes of \cite{mahajan2018fine}, as well as {\em gender} denoted as $G_2$ ($G_2=1$ for Female). The EHR surrogates $\bX$ include $X_1$, the log-transformed total count of the International Classification of Diseases (ICD) codes for T2D and $X_2$, the value of hemoglobin A1C obtained via laboratory tests. In addition, we have collected $Y^*$ on a subset of $n=269$ patients as the manual chart reviewing label for T2D status created by clinicians in 2014. Due to the gap of time windows of data collection, $Y^*$ is an imperfect label for the true T2D status $Y$ with its potential measurement error coming from the missingness of information between 2014 and 2021, as well as the switch of the ICD system from version 9 to 10 around 2015 at MGB. For the purpose of validation, we also extract the chart reviewing labels created by clinicians according to all information up to 2021 on a random subsample of the data with size $n_v=220$. These labels are more close to (arguably identical to) the true T2D status $Y$ and only used for validation and evaluation of the estimators trained on the set $\Oscr = \{\bO_i = (Y^*_i\delta_i, \delta_i, \bX_i, \G_i): i = 1,2, ...,N\}$.

In addition to Hong et al. 2019 and Naive-Logistic studied in Section \ref{sec:simu}, we also include four simple benchmark estimators including those obtained through the logistic regression against $\bG$ respectively using I(ICD$\geq$1), I(ICD$\geq$2), I(A1C$\geq $5.7) and I(A1C$\geq $6.4) as the binary outcomes. All of them are common and {\em convenient} ways to screen the subject with T2D frequently used in existing biomedical studies and practice. As the secondary analysis, we also estimate the AUC of the two important surrogates ICD and A1C using the imputation for $Y$ in TUBE and other methods except the aforementioned approaches directly using ICD or A1C to construct the outcome. This aim is slightly different from evaluating the derived phenotyping score $\alphahat(\bX)$ considered in Sections \ref{sec:method} and \ref{sec:simu} but it can be realized using nearly the same strategy and is typically more useful for clinicians and researchers in practice. We use 200 times bootstrap sampling to quantify the variance of all the estimators. The resulted estimators with their standard errors are presented in Table \ref{tab:real:1}.

Using the validation set with the true label $Y$, we obtain a validation estimator $\widehat\bbeta_v$ and evaluate the AUC of ICD and A1C. Evaluation metrics of the estimators for $\beta$ include: (1) mean square prediction error (MSPE) defined as the sample mean of $\{g(\bG_i\trans\bbetahat_v)-g(\bG_i\trans\bbetahat)\}^2$; (2) Deviance of the logistic model evaluated on the target data; (3) classifier's correlation (Class. Cor) with $\widehat\bbeta_v$, i.e., the sample correlation of $I(g(\bG_i\trans\bbetahat_v)>c)$ and $I(g(\bG_i\trans\bbetahat)>c)$ where $c$ is the sample mean of $g(\bG_i\trans\bbetahat_v)$; and (4) false classification rate (False Class.) compared to $\widehat\bbeta_v$, i.e., the empirical probability of $I(g(\bG_i\trans\bbetahat_v)>c)\neq I(g(\bG_i\trans\bbetahat)>c)$. The evaluation results are presented in Table \ref{tab:real:2}.

\begin{table}[htb!]
\centering
\begin{tabular}{cccccc}
  \hline
 & $\beta_0$ (Intercept) & $\beta_1$ (GRS) & $\beta_2$ (Gender) & AUC(ICD) & AUC(A1C) \\ 
  \hline
ICD$\geq 1$ & $-0.955_{0.028}$ & $0.649_{0.08}$ & $-0.556_{0.036}$ & -- & -- \\ 
ICD$\geq 2$  & $-1.286_{0.031}$ & $0.795_{0.087}$ & $-0.627_{0.04}$ & -- & -- \\ 
A1C$\geq 5.7$ & $-0.737_{0.027}$ & $0.464_{0.076}$ & $-0.461_{0.034}$ & -- & -- \\ 
A1C$\geq 6.5$ & $-2.1_{0.041}$ & $0.818_{0.115}$ & $-0.618_{0.053}$ & -- & -- \\ 
Naive-Logistic & $-1.386_{0.31}$ & $2.221_{0.639}$ & $-1.572_{0.377}$ & $0.949_{0.016}$ & $0.805_{0.023}$ \\ 
  Hong et al. 2019 & $-1.223_{0.136}$ & $1.204_{0.160}$ & $-0.806_{0.107}$ & $0.856_{0.046}$ & $0.787_{0.035}$ \\ 
  TUBE & $-1.352_{0.215}$ & $1.162_{0.200}$ & $-0.844_{0.140}$ & $0.973_{0.016}$ & $0.894_{0.013}$ \\ 
  Validation & $-1.341_{0.263}$ & $1.007_{0.854}$ & $-0.979_{0.387}$ & $0.983_{0.008}$ & $0.872_{0.036}$ \\ 
   \hline
\end{tabular}
\caption{\label{tab:real:1} Estimators for the T2D genetic model coefficient $\bbeta$ and the AUCs of ICD and A1C, with their empirical standard errors presented as subscriptions.}
\end{table}

\begin{table}[htb!]
\centering
\begin{tabular}{ccccc}
  \hline
 & MSPE & Deviance & Class. Cor & False Class. \\ 
  \hline
ICD$\geq 1$ & $0.0064$ & $0.004$ & $0.20$ & $0.46$ \\ 
ICD$\geq 2$ &  $0.0008$ & $-0.014$ & $0.81$ & $0.10$ \\ 
A1C$\geq 5.7$ & $0.0156$ & $0.029$ & $0$ & $0.50$ \\
A1C$\geq 6.4$  & $0.0069$ & $0.010$ & $0.12$ & $0.48$ \\ 
Naive-Logistic & $0.0034$ & $0.000$ & $0.40$ & $0.36$ \\ 
  Hong et al. 2019 & $0.0011$ & $-0.013$ & $0.81$ & $0.10$ \\ 
  TUBE & {$\mathbf{0.0002}$} & $\mathbf{-0.017}$ & $\mathbf{0.95}$ & $\mathbf{0.03}$ \\ 
  Validation & $0$ & $-0.017$ & $1$ & $0$ \\ 
   \hline
\end{tabular}
\caption{\label{tab:real:2} Estimation performance in the T2D genetic model $\bbeta$ evaluated using the metrics introduced in Section \ref{sec:real}.}
\end{table}

Among all methods under comparison, TUBE attains the closest point estimates to the validation estimator in terms of both 
$\bbeta$ and AUC. For example, the AUC of A1C evaluated using TUBE-imputed outcomes only differs from the the validation estimator by around $0.02$ while all the other estimators show more than $0.06$ gaps to the validation estimator. The estimation performance in $\bbeta$ are depicted more carefully in Table \ref{tab:real:2} where TUBE achieves the best on all metrics among all estimators except for $\widehat\bbeta_v$. For example, compared to the recent method proposed by \cite{hong2019semi}, our method attains more than $70\%$ reduction on MSPE, and $0.14$ larger classifier's correlation with the validation estimator. These results illustrate the effectiveness of leveraging our semiparametric modeling strategy to reduce potential bias due to misspecification. Meanwhile, although TUBE involves more complicated nonparametric regression, it does not result in significant inflation of the standard errors compared to \cite{hong2019semi}, which is a benefit of using parametric regression (projection) in Stage III.

Our estimator of $\bbeta$ reveals that the GRS has a significant positive effect (log(OR)=$1.16$, 95\% CI: $[0.77,1.55]$) on the risk of T2D and men have significantly higher risk to develop T2D than women in our study cohort. Interestingly, the effect sizes estimated using the four simple EHR outcomes, i.e., I(ICD$\geq$1), I(ICD$\geq$2), I(A1C$\geq $5.7), and I(A1C$\geq $6.4) are all smaller than $\beta_1$ and $\beta_2$ estimated by TUBE. As an explanation of this observation, after we convert the error-prone EHR outcomes to binary variables, they will have the same scale as the true outcome $Y$ and, thus, showing weaker association with the risk factors than $Y$ due to their measurement errors. This can be justified under the key assumption that $\mbox{ICD, A1C}$ are independent with the baseline risk factors given the True T2D status.

\section{Discussion}

In summary, we propose TUBE, a novel unsupervised method for analyzing multiple error-prone EHR outcomes and noisy labels against baseline risk factors, such as genetic variants extracted from EHR linked biobanks. TUBE incorporates a nonparametric composite regression step, and then uses it to combine the EHR outcomes for phenotyping and derive a parametric genetic risk model through projection. Compared to existing methods, our semiparametric strategy has two advantages. First, the nonparametric composite construction at the first stage safeguards the unsupervised learning against potential bias due to model misspecification. Second, the derived parametric genetic risk model obtained through projection enhances interpretability and achieves and significantly reduced variance in comparison to a fully nonparametric approach. These advantages are supported by our comprehensive asymptotic analysis, simulations, and a real-world study.

We acknowledges several limitations and potential extensions of our work. First, the validity of our method is prone to severe violation of the conditional independence assumption between the EHR outcomes and the baseline covariates. This issue can be alleviated by incorporating (small) samples with the true labels to calibrate the unsupervised estimator derived from surrogates. Recent advancements in surrogate-assisted semi-supervised learning \citep{zhang2022prior,hou2023surrogate} are particularly relevant to this discussion. Second, our current setup focuses on binary disease status. In current biomedical studies, time to the onset of clinical events (e.g., cancer relapse) is often not readily available with their EHR surrogates subject to measurement errors. Simple estimates of the event time based on billing or procedure codes may poorly approximate the true outcome and lead to bias. Therefore, expanding TUBE to incorporate multiple sources of imperfect and temporal endpoints under the survival setting is a potential direction for future research. In addition, our current method only accommodates low-dimensional genetic variants and a single disease or phenotype. Recent large scale genome??? and phenome???wide studies \citep[e.g.]{huang2019gwas,verma2023diversity} provides a strong motivation for its extensions to accommodate high-dimensional or machine learning estimates of the genetic risk models and multi-phenotype studies.

\bibliographystyle{apalike} 
\bibliography{ref}

\newpage
\appendix

\setcounter{equation}{0}
\setcounter{algorithm}{0}
\setcounter{theorem}{0}
\setcounter{table}{0}
\setcounter{figure}{0}
\renewcommand{\thelemma}{A\arabic{lemma}}
\renewcommand{\theequation}{A\arabic{equation}}
\renewcommand{\thetable}{A\arabic{table}}
\renewcommand{\thefigure}{A\arabic{figure}}
\renewcommand{\theremark}{A\arabic{remark}}
\renewcommand{\thealgorithm}{A\arabic{algorithm}}

\section*{Appendix}

\section{Additional implementation details}\label{sec:app:impl}

\begin{algorithm}[H]
\caption{\label{alg:app:2} EM algorithm for maximizing the non-parametric log-likelihood function (\ref{equ:prof:llh}).}
{\bf Input:} Observed data $\Oscr = \{\bO_i = (\Naive-Logistic_i\delta_i, \delta_i, \bX_i, \G_i): i = 1,2, ...,N\}$, and the phenotyping score $\alphahat(\bx)$ derived in Algorithm \ref{alg:1}.
 ~\\
{\bf Initialize} with $\boldsymbol{\widetilde{\eta}}_{\alphahat}^{(0)} = \{\widetilde\Tsc\subalphaty^{(0)}(\cdot),\widetilde\blambda^{(0)}, \widetilde\bxi^{(0)}:y=0,1\}$ introduced in Algorithm \ref{alg:app:1}. Iterate on the following two steps for $r=0,1,\ldots,R$ until convergence.
 ~\\
{\bf E-step}. For each subject $i$, impute the probability for $Y_i$ conditional on $Y^*_i$ (if observed) or $\alphahat(\bX_i)$:
\[
\widetilde Y_{i0}^{(r+1)}=\delta_i\times\frac{\widetilde\lambda_{1Y_{i}^*}^{(r)}g_1\{\bpsi\trans(\bG_i)\widetilde\bxi^{(r)}\}}{\sum_{y=0}^1 \widetilde\lambda_{yY_{i}^*}^{(r)}g_y\{\bpsi\trans(\bG_i)\widetilde\bxi^{(r)}\}};\quad\widetilde Y_{i1}^{(r+1)}=\frac{-\nabla\widetilde\Tsc^{(r)}\subalphatone\{\alphahat(\bX_i)\}g_1\{\bpsi\trans(\bG_i)\widetilde\bxi^{(r)}\}}{-\sum_{y=0}^1\nabla\widetilde\Tsc^{(r)}\subalphaty\{\alphahat(\bX_i)\}g_y\{\bpsi\trans(\bG_i)\widetilde\bxi^{(r)}\}}.
\]
 ~\\
{\bf M-step}. Update $\boldsymbol{\eta}_{\alphahat}$ through the MLE specified with the imputed outcomes from the E-step: 
\begin{align*}
&\widetilde\lambda_{yk}^{(r+1)}= \frac{\sum_{i=1}^n  I (Y^*_{i}= k) \{\Ytilde_{i0}^{(r+1)}\}^y \{1-\Ytilde_{i0}^{(r+1)}\}^{1-y}} {\sum_{i=1}^n\{\Ytilde_{i0}^{(r+1)}\}^y \{1-\Ytilde_{i0}^{(r+1)}\}^{1-y}};\quad k=0,1,\ldots,K\\
&\widetilde\bxi^{(r+1)}=\argmax{\bxi}\sum_{i=1}^n \ell\left(\widetilde Y_{i0}^{(r+1)},\bpsi\trans(\bG_i)\bxi\right)+ \sumiN\ell\left(\widetilde Y_{i1}^{(r+1)}, \bpsi\trans(\bG_i)\bxi\right);\\
&\widetilde\Tsc^{(r)}\subalphaty(c)=\frac{\sum_{i=1}^N  I (\alphahat(\bX_i)> c) \{\Ytilde_{i1}^{(r+1)}\}^y \{1-\Ytilde_{i1}^{(r+1)}\}^{1-y}} {\sum_{i=1}^N\{\Ytilde_{i1}^{(r+1)}\}^y \{1-\Ytilde_{i1}^{(r+1)}\}^{1-y}},\quad y=0,1.
\end{align*}
 ~\\
{\bf Output:} The imputed outcomes $\Ytilde_{i0}=\widetilde Y_{i0}^{(R)}$ (if $\delta_i=1$) and $\Ytilde_{i1}=\widetilde Y_{i1}^{(R)}$ for $i=1,2,\ldots,N$.

\end{algorithm}

\begin{algorithm}[H]
\caption{\label{alg:app:1} Initialization of the EM Algorithms.}

{\bf For Algorithm \ref{alg:1}}, we define $Y^{\dagger}_i=I(\Naive-Logistic_i=1)$ for subjects $i=1,2,\ldots,n$ and obtain the initial estimators $\widehat\bxi^{(0)},\widehat\bzeta^{(0)},\widehat\mu^{(0)}$ through MLE:
\[
\widehat\mu^{(0)}=\frac{1}{n}\sumin Y^{\dagger}_i;\quad\widehat\bxi^{(0)}=\argmax{\bxi}\sum_{i=1}^n \ell\left(Y^{\dagger}_i,\bpsi\trans(\bG_i)\bxi\right);\quad\widehat\bzeta_j^{(0)}=\argmax{\bzeta_j}\sumin \ell\left(Y^{\dagger}_i, \bvarphi\trans_j(X_{ij})\bzeta_j\right).
\]
For $\widehat\blambda^{(0)}$, we set $\widetilde\lambda_{1K}^{(0)}=0.85$; $\widetilde\lambda_{1k}^{(0)}=0.15/K$ for $k=0,1,\ldots,K-1$ and $\widetilde\lambda_{00}^{(0)}=0.85$; $\widetilde\lambda_{0k}^{(0)}=0.15/K$ for $k=1,\ldots,K$, in the belief that $\Naive-Logistic$ is reliable.
~\\
{\bf For Algorithm \ref{alg:app:1}}, we set $\widetilde\blambda^{(0)}=\widehat\blambda$ and $ \widetilde\bxi^{(0)}=\widehat\bxi$ based on the results in Algorithm \ref{alg:1}, and take
\[
\widetilde\Tsc\subalphaty^{(0)}(c)=\frac{\sum_{i=1}^n  I (\alphahat(\bX_i)> c) I(Y^{\dagger}_i=y)} {\sum_{i=1}^nI(Y^{\dagger}_i=y)},\quad y=0,1.
\]

\end{algorithm}

\section{Additional numerical results}\label{sec:app:sim}
In this section, we attach more complete simulation results as a supplement to the main results presented in Section \ref{sec:simu}.

\begin{sidewaystable}[ht]
{\small
\centering
\caption{Biases of parameter estimates over 500 simulations for the regression parameters for genetic effects ($\bbeta$), the area under the curve (AUC) for the classification algorithm, and the errors and/or uncertainties in labels ($\blambda$) for settings (a) with linear genetic effects, (b) with nonlinear genetic effects, and (c) with nonlinear genetic effects and slight violation of conditional independence between $\G$ and $\bX$.}
\resizebox{\textwidth}{!}{
\begin{tabular}{ccccccccccccc}
  (a)\\
  \hline
Method & $\beta_0$=-4.600 & $\beta_1$= 1.600 & $\beta_2$= 1.600 & $\beta_3$= 1.600 & $\beta_4$= 1.600 & AUC=0.702 & $\lambda_1(0)$=0.320 & $\lambda_1(0.5)$=0.490 & $\lambda_1(1)$=0.190 & $\lambda_0(0)$=0.700 & $\lambda_0(0.5)$=0.280 & $\lambda_0(1)$=0.030 \\ 
  \hline
Naive-Logistic$_{100}$ &  2.965 & -1.354 & -1.351 & -1.343 & -1.328 & -0.088 &   - &   - &   - &   - &   - &   - \\ 
  Hong et al$_{100}$ & -0.050 &  0.013 &  0.006 &  0.020 &  0.014 &  0.001 &  0.004 &  0.000 & -0.004 & -0.003 & -0.002 &  0.005 \\
  TUBE$_{100}$ & -0.145 &  0.040 &  0.036 &  0.049 &  0.044 & -0.001 &  0.001 &  0.002 & -0.003 &  0.002 & -0.005 &  0.003 \\
  Naive-Logistic$_{500}$ &  3.024 & -1.358 & -1.348 & -1.357 & -1.344 & -0.103 &   - &   - &   - &   - &   - &   - \\ 
  Hong et al$_{500}$ & -0.011 &  0.004 &  0.000 &  0.007 &  0.004 &  0.002 & -0.001 &  0.001 &  0.001 &  0.001 & -0.001 &  0.000 \\
  TUBE$_{500}$ & -0.089 &  0.029 &  0.025 &  0.029 &  0.031 &  0.000 & -0.004 &  0.002 &  0.002 &  0.007 & -0.003 & -0.003 \\
  Naive-Logistic$_{1000}$ &  3.019 & -1.360 & -1.346 & -1.347 & -1.349 & -0.104 &   - &   - &   - &   - &   - &   - \\ 
  Hong et al$_{1000}$ & -0.010 &  0.000 & -0.002 &  0.005 &  0.000 &  0.002 & -0.003 &  0.001 &  0.002 &  0.003 & -0.002 & -0.001 \\
  TUBE$_{1000}$ & -0.073 &  0.020 &  0.020 &  0.026 &  0.022 &  0.000 & -0.006 &  0.003 &  0.003 &  0.008 & -0.005 & -0.004 \\
   \hline
\end{tabular}
}
\resizebox{\textwidth}{!}{
\begin{tabular}{ccccccccccccc}
 (b)\\
    \hline
Method & $\beta_0$= 1.300 & $\beta_1$= 0.700 & $\beta_2$=-0.700 & $\beta_3$=-0.700 & $\beta_4$=-0.700 & AUC=0.702 & $\lambda_1(0)$=0.320 & $\lambda_1(0.5)$=0.490 & $\lambda_1(1)$=0.190 & $\lambda_0(0)$=0.700 & $\lambda_0(0.5)$=0.280 & $\lambda_0(1)$=0.030 \\ 
  \hline
Naive-Logistic$_{100}$ & -1.879 & -0.514 &  0.472 &  0.497 &  0.496 & -0.080 &   - &   - &   - &   - &   - &   - \\ 
  Hong et al$_{100}$ & -1.255 &  3.467 & -0.642 & -0.646 & -0.630 & -0.027 &  0.016 & -0.011 & -0.005 & -0.056 &  0.031 &  0.025 \\
  TUBE$_{100}$ &  0.020 &  0.016 & -0.024 & -0.009 & -0.011 & -0.003 & -0.001 &  0.003 & -0.002 &  0.003 & -0.004 &  0.001 \\ 
  Naive-Logistic$_{500}$ & -1.853 & -0.507 &  0.495 &  0.490 &  0.500 & -0.097 &   - &   - &   - &   - &   - &   - \\ 
  Hong et al$_{500}$ & -1.272 &  3.513 & -0.648 & -0.654 & -0.644 & -0.028 &  0.010 & -0.006 & -0.004 & -0.059 &  0.033 &  0.026 \\
  TUBE$_{500}$ &  0.011 &  0.013 & -0.017 & -0.007 & -0.004 & -0.001 & -0.002 &  0.000 &  0.002 &  0.000 &  0.000 &  0.000 \\ 
  Naive-Logistic$_{1000}$ & -1.850 & -0.509 &  0.495 &  0.493 &  0.500 & -0.097 &   - &   - &   - &   - &   - &   - \\ 
  Hong et al$_{1000}$ & -1.281 &  3.524 & -0.650 & -0.652 & -0.643 & -0.028 &  0.008 & -0.002 & -0.005 & -0.060 &  0.033 &  0.027 \\
  TUBE$_{1000}$ &  0.004 &  0.008 & -0.012 & -0.003 & -0.002 & -0.001 & -0.007 &  0.004 &  0.003 &  0.001 & -0.001 &  0.000 \\
   \hline
\end{tabular}
}
\resizebox{\textwidth}{!}{
\begin{tabular}{ccccccccccccc}
 (c)\\
     \hline
Method & $\beta_0$= 1.300 & $\beta_1$=-0.300 & $\beta_2$=-0.700 & $\beta_3$=-0.700 & $\beta_4$=-0.800 & AUC=0.702 & $\lambda_1(0)$=0.320 & $\lambda_1(0.5)$=0.490 & $\lambda_1(1)$=0.190 & $\lambda_0(0)$=0.700 & $\lambda_0(0.5)$=0.280 & $\lambda_0(1)$=0.030 \\ 
  \hline
Naive-Logistic$_{100}$ & -1.925 &  0.239 &  0.570 &  0.550 &  0.567 & -0.083 &     - &     - &     - &     - &     -&     - \\ 
  Hong et al$_{100}$ & -0.228 & -1.200 & -0.380 & -0.408 & -0.393 & -0.025 &  0.037 & -0.038 &  0.000 & -0.039 &  0.026 &  0.012 \\ 
  TUBE$_{100}$ &  0.090 &  0.046 & -0.021 & -0.032 & -0.029 & -0.009 &  0.017 & -0.007 & -0.010 & -0.007 &  0.004 &  0.002 \\ 
  Naive-Logistic$_{500}$ & -1.887 &  0.227 &  0.564 &  0.562 &  0.563 & -0.100 &     - &     - &     - &     - &     -&     -\\ 
  Hong et al$_{500}$ & -0.366 & -1.337 & -0.386 & -0.391 & -0.399 & -0.024 &  0.012 & -0.010 & -0.002 & -0.031 &  0.017 &  0.014 \\ 
  TUBE$_{500}$ &  0.065 &  0.044 & -0.013 & -0.023 & -0.019 & -0.003 & -0.008 &  0.003 &  0.005 &  0.005 & -0.003 & -0.001 \\
  Naive-Logistic$_{1000}$ & -1.887 &  0.226 &  0.557 &  0.568 &  0.571 & -0.102 &     - &     - &     - &     - &     -&     -\\ 
  Hong et al$_{1000}$ & -0.340 & -1.476 & -0.452 & -0.446 & -0.456 & -0.025 &  0.012 & -0.009 & -0.003 & -0.035 &  0.020 &  0.015 \\ 
  TUBE$_{1000}$ &  0.060 &  0.037 & -0.017 & -0.019 & -0.014 & -0.003 & -0.003 & -0.001 &  0.004 &  0.002 & -0.001 & -0.001 \\
   \hline
\end{tabular}
}
}

\end{sidewaystable}

\begin{sidewaystable}[ht]

\centering
\caption{Mean square errors (MSE) of parameter estimates over 500 simulations for the regression parameters for genetic effects ($\bbeta$), the area under the curve (AUC) for the classification algorithm, and the errors and/or uncertainties in labels ($\blambda$) for settings (a) with linear genetic effects, (b) with nonlinear genetic effects, and (c) with nonlinear genetic effects and slight violation of conditional independence between $\G$ and $\bX$.}

{\small
\centering
\resizebox{\textwidth}{!}{
\begin{tabular}{ccccccccccccc}
  (a)\\
  \hline
Method & $\beta_0$=-4.600 & $\beta_1$= 1.600 & $\beta_2$= 1.600 & $\beta_3$= 1.600 & $\beta_4$= 1.600 & AUC=0.702 & $\lambda_1(0)$=0.320 & $\lambda_1(0.5)$=0.490 & $\lambda_1(1)$=0.190 & $\lambda_0(0)$=0.700 & $\lambda_0(0.5)$=0.280 & $\lambda_0(1)$=0.030 \\ 
  \hline
  Naive-Logistic$_{100}$ & 9.092 & 1.863 & 1.885 & 1.870 & 1.826 & 0.009 &   - &   - &   - &   - &   - &   - \\ 
  Hong et al$_{100}$ & 0.678 & 0.070 & 0.075 & 0.085 & 0.072 & 0.000 & 0.005 & 0.005 & 0.003 & 0.012 & 0.011 & 0.002 \\
  TUBE$_{100}$ & 0.780 & 0.078 & 0.090 & 0.098 & 0.087 & 0.001 & 0.005 & 0.005 & 0.003 & 0.013 & 0.011 & 0.002 \\ 
  Naive-Logistic$_{500}$ & 9.194 & 1.849 & 1.830 & 1.851 & 1.815 & 0.011 &   - &   - &   - &   - &   - &   - \\ 
  Hong et al$_{500}$ & 0.620 & 0.064 & 0.070 & 0.077 & 0.065 & 0.000 & 0.001 & 0.001 & 0.001 & 0.003 & 0.002 & 0.000 \\ 
  TUBE$_{500}$ & 0.670 & 0.068 & 0.077 & 0.082 & 0.073 & 0.000 & 0.001 & 0.001 & 0.001 & 0.003 & 0.002 & 0.001 \\ 
  Naive-Logistic$_{1000}$ & 9.137 & 1.852 & 1.816 & 1.819 & 1.825 & 0.011 &   - &   - &   - &   - &   - &   - \\ 
  Hong et al$_{1000}$ & 0.604 & 0.060 & 0.065 & 0.076 & 0.066 & 0.000 & 0.001 & 0.001 & 0.000 & 0.001 & 0.001 & 0.000 \\ 
  TUBE$_{1000}$ & 0.660 & 0.064 & 0.072 & 0.079 & 0.076 & 0.000 & 0.001 & 0.001 & 0.000 & 0.002 & 0.001 & 0.000 \\
   \hline
\end{tabular}
}
\resizebox{\textwidth}{!}{
\begin{tabular}{ccccccccccccc}
  (b)\\
    \hline
Method & $\beta_0$= 1.300 & $\beta_1$= 0.700 & $\beta_2$=-0.700 & $\beta_3$=-0.700 & $\beta_4$=-0.700 & AUC=0.702 & $\lambda_1(0)$=0.320 & $\lambda_1(0.5)$=0.490 & $\lambda_1(1)$=0.190 & $\lambda_0(0)$=0.700 & $\lambda_0(0.5)$=0.280 & $\lambda_0(1)$=0.030 \\ 
  \hline
Naive-Logistic$_{100}$ &  3.896 &  0.311 &  0.302 &  0.320 &  0.321 &  0.008 &    - &    - &    - &    - &    - &    - \\ 
  Hong et al$_{100}$ &  2.145 & 13.794 &  0.564 &  0.615 &  0.557 &  0.001 &  0.023 &  0.024 &  0.014 &  0.006 &  0.004 &  0.001 \\ 
  TUBE$_{100}$ &  0.081 &  0.014 &  0.015 &  0.016 &  0.015 &  0.001 &  0.017 &  0.019 &  0.009 &  0.005 &  0.005 &  0.001 \\
  Naive-Logistic$_{500}$ &  3.489 &  0.265 &  0.259 &  0.253 &  0.262 &  0.010 &    - &    - &    - &    - &    - &    - \\ 
  Hong et al$_{500}$ &  2.191 & 13.569 &  0.533 &  0.574 &  0.548 &  0.001 &  0.004 &  0.004 &  0.002 &  0.004 &  0.002 &  0.001 \\ 
  TUBE$_{500}$ &  0.075 &  0.013 &  0.015 &  0.014 &  0.014 &  0.000 &  0.004 &  0.004 &  0.002 &  0.001 &  0.001 &  0.000 \\
  Naive-Logistic$_{1000}$ &  3.451 &  0.264 &  0.251 &  0.249 &  0.256 &  0.010 &    - &    - &    - &    - &    - &    - \\ 
  Hong et al$_{1000}$ &  2.176 & 13.639 &  0.539 &  0.567 &  0.541 &  0.001 &  0.002 &  0.002 &  0.001 &  0.004 &  0.001 &  0.001 \\ 
  TUBE$_{1000}$ &  0.065 &  0.011 &  0.013 &  0.012 &  0.012 &  0.000 &  0.002 &  0.002 &  0.001 &  0.001 &  0.000 &  0.000 \\
   \hline
\end{tabular}
}
\resizebox{\textwidth}{!}{
\begin{tabular}{ccccccccccccc}
  (c)\\
   \hline
Method & $\beta_0$= 1.300 & $\beta_1$=-0.300 & $\beta_2$=-0.700 & $\beta_3$=-0.700 & $\beta_4$=-0.800 & AUC=0.702 & $\lambda_1(0)$=0.320 & $\lambda_1(0.5)$=0.490 & $\lambda_1(1)$=0.190 & $\lambda_0(0)$=0.700 & $\lambda_0(0.5)$=0.280 & $\lambda_0(1)$=0.030 \\ 
  \hline
Naive-Logistic$_{100}$ & 4.043 & 0.105 & 0.398 & 0.378 & 0.390 &    - &    - &    - &    - &    - &    - \\ 
  Hong et al$_{100}$ & 2.667 & 4.855 & 0.360 & 0.536 & 0.440 & 0.001 & 0.050 & 0.052 & 0.022 & 0.007 & 0.005 & 0.001 \\ 
  TUBE$_{100}$ & 0.135 & 0.016 & 0.023 & 0.023 & 0.025 & 0.003 & 0.028 & 0.029 & 0.011 & 0.005 & 0.005 & 0.001 \\ 
  Naive-Logistic$_{500}$ & 3.629 & 0.060 & 0.332 & 0.330 & 0.331 & 0.010 &    - &    - &    - &    - &    - &    - \\ 
  Hong et al$_{500}$ & 2.297 & 4.564 & 0.344 & 0.335 & 0.359 & 0.001 & 0.012 & 0.010 & 0.004 & 0.003 & 0.001 & 0.001 \\ 
  TUBE$_{500}$ & 0.130 & 0.013 & 0.022 & 0.021 & 0.023 & 0.000 & 0.006 & 0.006 & 0.003 & 0.001 & 0.001 & 0.000 \\ 
  Naive-Logistic$_{1000}$ & 3.589 & 0.055 & 0.317 & 0.328 & 0.333 & 0.011 &    - &    - &    - &    - &    - &    - \\ 
  Hong et al$_{1000}$ & 4.793 & 7.153 & 0.912 & 1.001 & 1.280 & 0.001 & 0.008 & 0.006 & 0.003 & 0.003 & 0.001 & 0.001 \\
  TUBE$_{1000}$ & 0.113 & 0.012 & 0.019 & 0.019 & 0.022 & 0.000 & 0.003 & 0.003 & 0.002 & 0.001 & 0.001 & 0.000 \\
   \hline
\end{tabular}
}
}
\end{sidewaystable}

\begin{sidewaystable}[ht]
\centering
\caption{Coverage probabilities (CP) at the 95\% nominal level of parameter estimates over 500 simulations for the regression parameters for genetic effects ($\bbeta$), the area under the curve (AUC) for the classification algorithm, and the errors and/or uncertainties in labels ($\blambda$) for settings (a) with linear genetic effects, (b) with nonlinear genetic effects, and (c) with nonlinear genetic effects and slight violation of conditional independence between $\G$ and $\bX$.}

{\small
\centering
\resizebox{\textwidth}{!}{
\begin{tabular}{ccccccccccccc}
  (a)\\
  \hline
Method & $\beta_0$=-4.600 & $\beta_1$= 1.600 & $\beta_2$= 1.600 & $\beta_3$= 1.600 & $\beta_4$= 1.600 & AUC=0.702 & $\lambda_1(0)$=0.320 & $\lambda_1(0.5)$=0.490 & $\lambda_1(1)$=0.190 & $\lambda_0(0)$=0.700 & $\lambda_0(0.5)$=0.280 & $\lambda_0(1)$=0.030 \\ 
  \hline
  Naive-Logistic$_{100}$ & 0.002 & 0.000 & 0.000 & 0.000 & 0.002 & 0.398 &  - &   - &   - &   - &   - &   - \\ 
  Hong et al$_{100}$ & 0.946 & 0.954 & 0.958 & 0.948 & 0.950 & 0.946 & 0.942 & 0.954 & 0.952 & 0.956 & 0.960 & 0.934 \\
  TUBE$_{100}$ & 0.940 & 0.944 & 0.944 & 0.940 & 0.938 & 0.998 & 0.952 & 0.958 & 0.950 & 0.954 & 0.958 & 0.936 \\ 
  Naive-Logistic$_{500}$ & 0.000 & 0.000 & 0.000 & 0.000 & 0.000 & 0.000 &   - &   - &   - &   - &   - &   - \\ 
  Hong et al$_{500}$ & 0.946 & 0.956 & 0.952 & 0.950 & 0.952 & 0.948 & 0.940 & 0.946 & 0.952 & 0.954 & 0.946 & 0.960 \\ 
  TUBE$_{500}$ & 0.942 & 0.946 & 0.946 & 0.948 & 0.954 & 0.954 & 0.948 & 0.948 & 0.944 & 0.948 & 0.946 & 0.972 \\ 
  Naive-Logistic$_{1000}$ & 0.000 & 0.000 & 0.000 & 0.000 & 0.000 & 0.000 &   - &   - &   - &   - &   - &  - \\ 
  Hong et al$_{1000}$ & 0.948 & 0.950 & 0.946 & 0.956 & 0.950 & 0.954 & 0.938 & 0.950 & 0.948 & 0.944 & 0.952 & 0.968 \\ 
  TUBE$_{1000}$ & 0.954 & 0.952 & 0.938 & 0.954 & 0.940 & 0.954 & 0.938 & 0.950 & 0.938 & 0.942 & 0.952 & 0.978 \\
   \hline
\end{tabular}
}
\resizebox{\textwidth}{!}{
\begin{tabular}{ccccccccccccc}
  (b)\\
 \hline
Method & $\beta_0$= 1.300 & $\beta_1$= 0.700 & $\beta_2$=-0.700 & $\beta_3$=-0.700 & $\beta_4$=-0.700 & AUC=0.702 & $\lambda_1(0)$=0.320 & $\lambda_1(0.5)$=0.490 & $\lambda_1(1)$=0.190 & $\lambda_0(0)$=0.700 & $\lambda_0(0.5)$=0.280 & $\lambda_0(1)$=0.030 \\ 
  \hline
Naive-Logistic$_{100}$ & 0.097 & 0.333 & 0.628 & 0.554 & 0.547 & 0.574 &   - &   - &  - &  - &  - &   - \\ 
  Hong et al$_{100}$ & 0.634 & 0.261 & 0.675 & 0.752 & 0.697 & 0.602 & 0.952 & 0.956 & 0.956 & 0.828 & 0.903 & 0.871 \\ 
  TUBE$_{100}$ & 0.947 & 0.956 & 0.929 & 0.945 & 0.958 & 0.992 & 0.952 & 0.954 & 0.941 & 0.949 & 0.952 & 0.947 \\ 
  Naive-Logistic$_{500}$ & 0.000 & 0.000 & 0.008 & 0.004 & 0.002 & 0.000 &   - &   - &  - &  - &  - &   - \\ 
  Hong et al$_{500}$ & 0.640 & 0.095 & 0.554 & 0.628 & 0.628 & 0.554 & 0.954 & 0.966 & 0.956 & 0.341 & 0.729 & 0.408 \\ 
  TUBE$_{500}$ & 0.958 & 0.952 & 0.952 & 0.958 & 0.964 & 0.941 & 0.954 & 0.956 & 0.943 & 0.943 & 0.947 & 0.927 \\ 
  Naive-Logistic$_{1000}$ & 0.000 & 0.000 & 0.000 & 0.000 & 0.000 & 0.000 &   - &   - &  - &  - &  - &   - \\ 
  Hong et al$_{1000}$ & 0.604 & 0.083 & 0.566 & 0.636 & 0.618 & 0.543 & 0.943 & 0.941 & 0.947 & 0.083 & 0.475 & 0.121 \\ 
  TUBE$_{1000}$ & 0.954 & 0.960 & 0.949 & 0.956 & 0.958 & 0.939 & 0.954 & 0.943 & 0.943 & 0.947 & 0.945 & 0.947 \\
   \hline
\end{tabular}
}
\resizebox{\textwidth}{!}{
\begin{tabular}{ccccccccccccc}
  (c)\\
   \hline
Method & $\beta_0$= 1.300 & $\beta_1$=-0.300 & $\beta_2$=-0.700 & $\beta_3$=-0.700 & $\beta_4$=-0.800 & AUC=0.702 & $\lambda_1(0)$=0.320 & $\lambda_1(0.5)$=0.490 & $\lambda_1(1)$=0.190 & $\lambda_0(0)$=0.700 & $\lambda_0(0.5)$=0.280 & $\lambda_0(1)$=0.030 \\ 
  \hline
Naive-Logistic$_{100}$ & 0.079 & 0.797 & 0.436 & 0.482 & 0.428 & 0.522 &   - &   - &   - &   - &   - &   -\\ 
  Hong et al$_{100}$ & 0.956 & 0.937 & 0.896 & 0.954 & 0.927 & 0.858 & 0.956 & 0.927 & 0.958 & 0.925 & 0.929 & 0.937 \\
  TUBE$_{100}$ & 0.939 & 0.935 & 0.960 & 0.948 & 0.935 & 0.985 & 0.971 & 0.952 & 0.969 & 0.956 & 0.954 & 0.952 \\ 
  Naive-Logistic$_{500}$ & 0.000 & 0.290 & 0.002 & 0.006 & 0.006 & 0.002 &   - &   - &   - &   - &   - &   -\\ 
  Hong et al$_{500}$ & 0.933 & 0.881 & 0.862 & 0.868 & 0.864 & 0.839 & 0.933 & 0.952 & 0.952 & 0.931 & 0.944 & 0.937 \\ 
  TUBE$_{500}$ & 0.950 & 0.929 & 0.950 & 0.939 & 0.942 & 0.942 & 0.946 & 0.952 & 0.950 & 0.946 & 0.944 & 0.927 \\ 
  Naive-Logistic$_{1000}$ & 0.000 & 0.077 & 0.000 & 0.000 & 0.000 & 0.000 &   - &   - &   - &   - &   - &   -\\ 
  Hong et al$_{1000}$ & 0.985 & 0.946 & 0.979 & 0.987 & 0.990 & 0.843 & 0.939 & 0.946 & 0.948 & 0.879 & 0.912 & 0.894 \\ 
  TUBE$_{1000}$ & 0.942 & 0.937 & 0.946 & 0.946 & 0.948 & 0.946 & 0.952 & 0.958 & 0.933 & 0.958 & 0.954 & 0.939 \\
   \hline
\end{tabular}
}
}
\end{sidewaystable}

\end{document}